\documentclass[a4paper,UKenglish,cleveref, autoref]{article}

\title{Building large $k$-cores from sparse graphs}

\author{
Fedor V. Fomin\thanks{
Department of Informatics, University of Bergen, Norway.}
\and
Danil Sagunov\thanks{
    St.\ Petersburg Department of V.A.\ Steklov Institute of Mathematics, Russia.
}\thanks{JetBrains Research, Russia}
\and 
Kirill Simonov\addtocounter{footnote}{-3}\footnotemark{}
}

\newif\ifshort
\shortfalse

\usepackage{url}
\usepackage{hyperref}
\usepackage{comment}
\usepackage{amsthm}
\usepackage[classfont=roman]{complexity}
\usepackage{amsmath}
\usepackage{newclude}
\usepackage{amsfonts,amssymb}
\usepackage{mathbbol}
\usepackage{boxedminipage}
\usepackage{framed}
\usepackage{thmtools}
\usepackage{thm-restate}
\usepackage{xspace}
\usepackage{todonotes}
\usepackage{xifthen}
\usepackage{tabularx}
\usepackage{tikz}
\usetikzlibrary{calc}

\makeatletter
\def\th@plain{%
	\thm@notefont{}
	\itshape 
}
\def\th@definition{%
	\thm@notefont{}
	\normalfont 
}
\makeatother

\newcommand{\pname}{\textsc}
\newcommand{\ProblemFormat}[1]{\pname{#1}}
\newcommand{\ProblemIndex}[1]{\index{problem!\ProblemFormat{#1}}}
\newcommand{\ProblemName}[1]{\ProblemFormat{#1}\ProblemIndex{#1}{}\xspace}

\newcommand{\probCore}{\ProblemName{Edge $k$-Core}}

\newlength{\RoundedBoxWidth}
\newsavebox{\GrayRoundedBox}
\newenvironment{GrayBox}[1]%
{\setlength{\RoundedBoxWidth}{.93\linewidth}
	\def\boxheading{#1}
	\begin{lrbox}{\GrayRoundedBox}
		\begin{minipage}{\RoundedBoxWidth}}%
		{   \end{minipage}
	\end{lrbox}
	\begin{center}
		\begin{tikzpicture}%
		\node(Text)[draw=black!20,fill=white,rounded corners,%
		inner sep=2ex,text width=\RoundedBoxWidth]%
		{\usebox{\GrayRoundedBox}};
		\coordinate(x) at (current bounding box.north west);
		\node [draw=white,rectangle,inner sep=3pt,anchor=north west,fill=white] 
		at ($(x)+(6pt,.75em)$) {\boxheading};
		\end{tikzpicture}
\end{center}}     

\newenvironment{defproblemx}[2][]{\noindent\ignorespaces%
	\FrameSep=6pt%
	\parindent=0pt%
	\vspace*{-1.5em}
	\ifthenelse{\isempty{#1}}{%
		\begin{GrayBox}{\textsc{#2}}%
		}{%
			\begin{GrayBox}{\textsc{#2} parameterized by~{#1}}%
			}
			\begin{tabular*}{\linewidth}{@{\hspace{.1em}} >{\itshape} p{1.8cm} p{0.8\linewidth} @{}}%
			}{
			\end{tabular*}%
		\end{GrayBox}%
		\ignorespacesafterend
	}  
	
	\newcommand{\defparproblema}[4]{
		\begin{defproblemx}[#3]{#1}
			Input:  & #2 \\
			Task: & #4
		\end{defproblemx}
	}%
	
	\newcommand{\defproblema}[3]{
		\begin{defproblemx}{#1}
			Input:  & #2 \\
			Task: & #3
		\end{defproblemx}
	}%
	
\renewcommand{\O}{\mathcal{O}}
\newcommand{\tw}{\mathrm{tw}}
\newcommand{\vc}{\mathrm{vc}}
\newcommand{\defc}{\operatorname{df}}

\usepackage{environ}

\ifshort
\makeatletter
\newenvironment{claimO}[1][]{
	\if\relax\detokenize{#1}\relax
	\expandafter\@firstoftwo
	\else
	\expandafter\@secondoftwo
	\fi
	{\begin{claim}[$\star$]}{\begin{claim}[#1, $\star$]}
		}
		{
		\end{claim}
	}
	\makeatother
\makeatletter
\newenvironment{lemmaO}[1][]{
	\if\relax\detokenize{#1}\relax
	\expandafter\@firstoftwo
	\else
	\expandafter\@secondoftwo
	\fi
	{\begin{lemma}[$\star$]}{\begin{lemma}[#1, $\star$]}
		}
		{
		\end{lemma}
	}
	\makeatother
\makeatletter
\newenvironment{propositionO}[1][]{
	\if\relax\detokenize{#1}\relax
	\expandafter\@firstoftwo
	\else
	\expandafter\@secondoftwo
	\fi
	{\begin{proposition}[$\star$]}{\begin{proposition}[#1, $\star$]}
		}
		{
		\end{proposition}
	}
	\makeatother
	\makeatletter
	\newenvironment{theoremO}[1][]{
		\if\relax\detokenize{#1}\relax
		\expandafter\@firstoftwo
		\else
		\expandafter\@secondoftwo
		\fi
		{\begin{theorem}[$\star$]}{\begin{theorem}[#1, $\star$]}
			}
			{
			\end{theorem}
		}
		\makeatother
		\makeatletter
		\newenvironment{corollaryO}[1][]{
			\if\relax\detokenize{#1}\relax
			\expandafter\@firstoftwo
			\else
			\expandafter\@secondoftwo
			\fi
			{\begin{corollary}[$\star$]}{\begin{corollary}[#1, $\star$]}
				}
				{
				\end{corollary}
			}
			\makeatother
			\NewEnviron{proofO}{}
			\else

			\makeatletter
			\newenvironment{proofO}[1][]{\begin{proof}[#1]}{\end{proof}}
			\makeatother
			
			\fi

\newcounter{inequnum}

\makeatletter
\newcommand*{\inlineequation}[2][]{%
	\begingroup
	\refstepcounter{inequnum}%
	\ifx\\#1\\%
	\else
	\label{#1}%
	\fi
	\relpenalty=10000 %
	\binoppenalty=10000 %
	~(\theinequnum)\,\,
	\ensuremath{%
		#2%
	}%

	\endgroup
}
\makeatother

\newcommand{\cO}{\mathcal{O}}
\newcommand{\Oh}{\cO}
\newcommand{\polyn}{n^{\Oh(1)}}
\newcommand{\probILP}{\ProblemName{Integer Linear Programming Feasibility}}
\newcommand{\probILPshort}{\ProblemName{ILP}}

\newtheorem{theorem}{Theorem}
\newtheorem{proposition}[theorem]{Proposition}
\newtheorem{lemma}[theorem]{Lemma}

\newtheorem{claim}[theorem]{Claim}
\newtheorem{corollary}{Corollary}
\theoremstyle{definition}
\newtheorem{definition}[theorem]{Definition}
\newenvironment{claimproof}[1]{\begin{proof}}{\end{proof}}

\begin{document}

\maketitle



\ifshort
\else
\maketitle
\fi
\begin{abstract}
A popular model to measure  network stability  is the  $k$-core, that is the maximal induced subgraph in which every vertex has degree at least $k$.  For example, $k$-cores are commonly used to model the unraveling phenomena in social networks. In this model, users having less than $k$ connections within the network leave it, so the remaining users form exactly the $k$-core. In this paper we study the question of whether  it is possible to make the network more robust by spending only a limited amount of resources on new connections. 
 A mathematical model for the $k$-core construction problem  is the following  \probCore optimization problem. We are given 
a graph $G$ and integers $k$,  $b$ and
$p$. The task is to ensure that the $k$-core of $G$ has at least $p$ vertices by adding at most $b$ edges. 

The previous studies on  \probCore demonstrate that the problem is computationally challenging. 
In particular, it is NP-hard when $k=3$,  W[1]-hard when parameterized by $k+b+p$ (Chitnis and Talmon,   2018), and  APX-hard (Zhou et al,  2019). 
 Nevertheless,  we show  that there are efficient algorithms with provable guarantee when the $k$-core has to be constructed from a sparse graph with some additional structural properties. 
Our results are
\begin{itemize}
\item When the input graph is a forest,   \probCore is solvable in polynomial time;
\item  \probCore is fixed-parameter tractable (FPT) when parameterized by the minimum size of a vertex 
cover in the input graph. On the other hand, with such parameterization, the problem does not admit a polynomial kernel subject to a widely-believed assumption from complexity theory;
\item    \probCore   is \FPT{} parameterized by the treewidth of the graph plus $k$. This improves upon a result of Chitnis and Talmon by not requiring $b$ to be small.
\end{itemize}
Each of our  algorithms is built upon a new graph-theoretical result interesting in its own. 
\end{abstract}

\section{Introduction}

The $k$-core in an undirected graph $G$ is the maximal induced subgraph of $G$ in which all vertices have degree at least $k$. This concept has been applied in various  areas including social networks \cite{BhawalkarKLRS15,chwe1999structure,chwe2000communication}, protein function prediction 
\cite{wuchty2005peeling},
 hierarchical structure analysis \cite{alvarez2008k},  graph visualization \cite{Alvarez-HamelinDBV05}, and network clustering  and connectivity \cite{Alvarez-HamelinBB11,giatsidis2014corecluster}. 


In online social networks users tend to contribute content only when a certain amount of their friends do the same  
\cite{DBLP:conf/chi/BurkeML09}, or in other words, when the formed community is a $k$-core for some threshold parameter $k$. Interestingly, losing even a small amount of users or links can bring to the 
  {cascade of iterated withdrawals}. 
A classical example of such phenomena  is the example of Schelling from \cite{schelling2006micromotives}: Consider a cycle on $n$ vertices, which  is a $2$-core with $n$ vertices. Missing just one edge from this graph turns it into in a path and   triggers the withdrawals that results in dismounting 
of the whole network.  On the other hand, adding a small number of extra links can create a large $k$-core and thus prevent users from withdrawal.  We consider the following mathematical model for this problem. 
For a given a network, the assumption is  that a user leaves the network when less than $k$ his/her friends remain within it.
We would like to prevent unraveling of the network, so that at least $p$ users remain engaged in it.
To achieve this, we are given a budget to establish at most $b$ new connections between the users of the network.
 More precisely, the problem is stated as follows. 

 \defproblema{\probCore}%
{A simple undirected graph $G$ and integers $b$, $k$, and $p$.}%
{Decide whether there exists $B\subseteq\binom{V(G)}{2}\setminus E(G)$ of size at most $b$ such that the $k$-core of the graph $(V(G), E(G) \cup B)$ is of size at least $p$.}


The \probCore problem was introduced by  Chitnis and Talmon  in \cite{Chitnis2018} as a model of 
preventing unraveling  
in   networks. 
For instance,  in a P2P network, any user benefiting from the network should be linked to at least $k$ other users exchanging resources. In this scenario the  \probCore  model could be used to find extra connections between users to provide a better service for larger number of users
  \cite{Chitnis2018,ZhouZLZ019}.
Other potential application of \probCore in
real-life networks include  friend recommendation in social networks, connection construction in telecom networks, etc. 
We find the \probCore problem to be interesting from the theoretical perspective too:  it has strong links to the well-studied family of problems, where one seeks a modification of  a graph satisfying certain conditions on vertex degrees, see \cite{DBLP:journals/corr/abs-2001-06867} for further references. 
Our interest in the study of the problem is of a theoretical nature.

   The $k$-core in a graph can be found by a simple “shaving” procedure: If a graph contains a vertex of degree less than $k$, then this vertex  cannot be in its $k$-core and thus can be safely removed. 
  Apparently,  solving \probCore is more challenging. In particular,  
  Chitnis and Talmon  in \cite{Chitnis2018} proved that  \probCore is NP-complete even for $k=3$ and when the input graph $G$ is $2$-degenerate.\footnote{Recall that a graph is $d$-degenerate if its every induced subgraph contains a vertex of degree at most $d$. Thus the $d$-core is the maximum subgraph which is not $d-1$ degenerate.} Moreover, the problem is W[1]-hard being parameterized by $k+b+p$. On the other hand, they show that if the treewidth of the graph $G$ is $\tw$, then the problem is solvable in time 
  $(k + \tw)^{\O(\tw + b)}\cdot n^{\O(1)}$ and hence is fixed-parameter tractable (FPT) parameterized by $k+\tw +b$. These results of Chitnis and Talmon are the departure point for our study.

 \smallskip\noindent\textbf{Our results.} 
 We study  the algorithmic complexity of  \probCore on three families of sparse graphs: forests, graphs with bounded vertex cover number and graphs of bounded treewidth.  Each of our algorithms is based on one of the common algorithmic paradigms:  dynamic programming  for forests and treewidth, and ILP for vertex cover. The interesting part here is that in each of the cases, the successful application of an algorithmic paradigm crucially depends on a  new combinatorial result.    
 We show the following.

 \smallskip\noindent\emph{Growing from forest.} 
 We prove (Theorem~\ref{thm:kcoreforest}) that   \probCore  is solvable in time  $\O(k \cdot |V(G)|^2)$, when the input graph $G$ is a forest.  The algorithm  is based on a dynamic programming over subtrees. The crucial part of the work is to make this algorithm run in polynomial time. For that we need a new graph-theoretical result, Theorem~\ref{thm:trees_optimal}.
 The theorem states that 
	for any integer $k$, a forest $F$ on at least $k+1$ vertices can be completed into a graph of minimum degree $k$ by adding at most 
		$$\Bigl\lceil\frac{1}{2}\sum\limits_{v \in V(F)}\max\{0, k-\deg(v)\}\Bigr\rceil$$
	edges. Moreover, this bound is tight, any forest requires such amount of edge additions to grow into a $k$-core. The proof of Theorem~\ref{thm:trees_optimal} is non-trivial and 
exploits an interesting connection between the cores in a graph and sufficient conditions on the  existence of a large matching in a graph. Here  the recent combinatorial theorem of 
 Henning and Yeo \cite{Henning2018}  on matchings in graphs of bounded degrees becomes handy. 
 
 \smallskip\noindent\emph{Bounded vertex cover.} We prove that the problem is \FPT{} parameterized by the minimum size of a vertex cover in a graph. More precisely, in Theorem~\ref{thm:vc:edgecore}, we give an 
 algorithm of running time $2^{\O(\operatorname{vc} \cdot 3^{\operatorname{vc}})}\cdot n^{\O(1)}$, where 
$\operatorname{vc}$ is the vertex cover  number of the input graph. Let us note that every graph is $\vc$-degenerate. We solve  the problem by reducing it to 
an  integer linear program (ILP), whose number of variables is bounded by some function of $\operatorname{vc}$. This allows to apply Lenstra's algorithm \cite{Lenstra1983}, see also
 \cite{Kannan1987,Frank1987}, to solve   \probCore. Nowadays ILP is a commonly used   tool for designing  parameterized algorithms, see   e.g. \cite[Chapter~6]{cygan2015parameterized}. 
 However, just like in the case of forests, the application of an  algorithmic paradigm is not direct. In order to encode the problem as ILP with the required number of variables, we need a  new combinatorial result  (Lemma~\ref{lemma:check_eg}) about degree sequences of a graph.  
 One of the components in the proof of Lemma~\ref{lemma:check_eg} is 
the classical Erd\H{o}s-Gallai theorem  \cite{erdHos1960grafok} about graphic sequences. 
We complement \FPT{} algorithm  by  lower bounds on the size of the kernel.

 \smallskip\noindent\emph{Bounded treewidth.}  
 Chitnis and Talmon  in \cite{Chitnis2018} have shown that   \probCore is \FPT{} parameterized by 
 $\tw+k+b$, where $\tw$ is the treewidth of the input graph. Even in the case when the treewidth and $k$ are constants, this does not mean that the problem is solvable in polynomial time. We enhance  this result by proving that   \probCore is \FPT{} parameterized by 
 $\tw+k$. As the algorithm of Chitnis and Talmon  in \cite{Chitnis2018}, our algorithm is a dynamic programming on graphs of bounded treewidth, but 
 again, in order to make it work, we need  a new combinatorial result (Theorem~\ref{lemma:largeopt}). When the budget $b$ is small (of order $k^3$), the algorithm of Chitnis and Talmon suffices. When 
 the budget $b$ is large, we are able to approach the problem in an interesting new way. Here  Theorem~\ref{lemma:largeopt}
provides us with a  criteria how  a subset of vertices can be turned into a $k$-core ``optimally''. 
  This key insight allows us to show that the problem is \FPT{} parameterized by  $\tw+k$.

   \smallskip\noindent\textbf{Related work.} 
   The usability of $k$-cores in the study of 
   network unraveling phenomena was popularized by the influential paper  of Bhawalkar et al.~\cite{BhawalkarKLRS12}  who suggested   the model of forcing a limited number of users of a network to stay in order to maximize the size of the $k$-core.
The same problem was further studied in \cite{ChitnisFG13}, where new computational results were obtained and some results of \cite{BhawalkarKLRS15} were strengthened.
Also, Chitnis, Fomin and Golovach studied this problem applied to networks where the underlying graph is directed \cite{DBLP:journals/iandc/ChitnisFG16}. Heuristic algorithms for this problem are discussed in \cite{zhang2017olak}.

\probCore was introduced in 
 \cite{Chitnis2018}, where also a number of complexity and algorithmic results about the problem were established. Zhou et al. \cite{ZhouZLZ019}
 provide some non-approximability results for  \probCore  as well as some heuristics. The work 
 \cite{zhang2017finding}  is devoted to the ``dual'' problem  of disengaging a limited number of users from a network in order to minimize its $k$-core size.
 Another work in this context is the work of Luo, Molter and Suchy~\cite{luo2018parameterized}.

More generally,  \probCore fits into a large class of edge modification problems, where one is seeking for an optimum modification to some desired graph property  \cite{DBLP:journals/corr/abs-2001-06867}.  
In particular, a significant part of literature in parameterized complexity is devoted to related problems of graph modification to graphs with some vertex degree properties like being regular, Euler, or to some degree sequence
~\cite{FominGPS16,Golovach14,Golovach14a,GoyalMPPS15,MathiesonS12}.

	\section{Preliminaries}\label{sec:prelim}
All graphs considered in this paper are simple undirected graphs.
We use standard graph notation and terminology, following the book of Diestel \cite{diestel2018graph}.
We write $G+F$ to denote the simple graph obtained by adding the edges from $F \subseteq \binom{V(G)}{2}\setminus E(G)$ to a graph $G$.
If not specified otherwise, we use $n$ to denote the number of vertices of the graph $G$ in an input instance of \probCore.

Throughout this paper, we use the following terms.
In the following definitions, we assume that $k$ is fixed.
\begin{definition}[Deficiency]
	For a graph $G$, and its vertex $v \in V(G)$, let $\defc_{G}(v)=\max\{0, k-\deg_G(v)\}$ denote the \emph{deficiency} of $v$ in $G$.
By $\defc(G)=\sum_{v\in V(G)} \defc_G(v)$ we denote the \emph{total deficiency} in $G$.
\end{definition}

Note that an addition of an edge between two vertices of $G$ can decrease $\defc(G)$ by at most two.
It also does not make any sense to add edges that do not decrease    deficiency if we aim to complete $G$ to a graph of minimum degree $k$. We distinguish added edges by whether they decrease deficiency by two or one.
\begin{definition}[Good/bad edges]
For nonadjacent vertices $u,v \in V(G)$ a new added edge $uv$ is  \emph{good} if both $\defc_G(u)>0$ and $\defc_G(v)>0$. 
If $\defc_G(u)=0$ and $\defc_G(v)>0$, then $uv$ is  \emph{bad}.
%
\end{definition}
Thus adding a good edge decreases the total deficiency by $2$ and adding a bad one by $1$.

\begin{definition}[A $k$-core graph]
    We say that a graph $G$ is \emph{a $k$-core} if $G$ is the $k$-core of itself. We also say that a vertex set $H$ in $G$ \emph{induces a $k$-core} in $G$ if $G[H]$ is a $k$-core.
\end{definition}
Note that whenever there is a vertex set $H$ of size at least $p$ which induces a $k$-core in $G$, the $k$-core of $G$ has also size at least $p$, since it is the unique maximal induced subgraph of $G$ which is a $k$-core.
We often use this simple observation throughout the paper whenever we show that the $k$-core is large by presenting a large vertex set which induces a $k$-core.

\section{Growing from forest}\label{sec_trees}
In this section we present our polynomial time algorithm for \probCore on forests and the underlying graph-theoretical result.

The algorithm itself is a dynamic programming over subtrees. Normally, an algorithm like this would go from leaves to larger and larger subtrees, storing for every subtree a list of possible configurations a solution could induce on this subtree.
In the \probCore problem, naturally we want to store information about edges added inside the subtree and vertices from the subtree which we may later connect to something outside.

Naively, this would take exponential space, as it seems we have to store at least the degrees of the selected vertices in the subtree.
However, the following theorem, which is the central technical result of this section, helps greatly.
\begin{theorem}\label{thm:trees_optimal}
	For any integer $k$, any forest $T$ on at least $k+1$ vertices can be completed to a graph of minimum degree $k$ by adding at most 
	$$\Bigl\lceil\frac{1}{2}\sum\limits_{v \in V(T)}\max\{0, k-\deg(v)\}\Bigr\rceil$$
	edges, and this  cannot be done with   less  edge additions.
    Moreover, in the case $k \ge 4$, it can be done in a way that the added edges form a connected graph on the vertices they cover.
\end{theorem}

For our algorithm, Theorem~\ref{thm:trees_optimal} means that whenever we fix the subset of vertices $H$, we have to add exactly $\lceil \defc(T[H]) /2 \rceil$ edges in order to induce a $k$-core on $H$.
Thus it is enough to find a subset of vertices $H$ of size at least $p$ with the smallest possible $\defc(T[H])$.
This objective turns out to be simple enough for the bottom-top dynamic programming. Namely, for a subtree $T_v$ rooted at $v$, it is enough to store the size of $H \cap T_v$, 
the total deficiency of these vertices, whether $v$ is in $H$ and how many neighbors in $H \cap T_v$ it has. Since $v$ separates $T_v$ from the rest of the tree, the deficiency of other vertices in $H \cap T_v$ is unchanged no matter how $H$ looks like in the rest of the tree.

The discussion above ultimately leads to a polynomial time algorithm, stated formally in the next theorem.

\begin{theoremO}\label{thm:kcoreforest}
	\probCore is solvable in time $\O(kn^2)$ on the class of forests.
\end{theoremO}
\ifshort
\else
\begin{proof}
	Let $(G, b, k, p)$ be the given instance of \textsc{Edge $k$-Core}, and $G$ be a forest.
	Consider a subset $H \subseteq V(G)$ of the vertices of $G$.
	Since $G[H]$ is a forest, by Theorem~\ref{thm:trees_optimal}, one needs $\lceil \frac{1}{2} \defc(G[H]) \rceil$ edge additions to make $H$ a (subset of) $k$-core in $G$.
	Thus, to solve the given instance, it is enough to check whether there is a subset $H \subseteq V(G)$ with $|H|\ge p$, such that $2b \ge \defc(G[H])$.
	Hence, it is enough to find $H$ with the smallest value of $\defc(G[H])$.
	In the rest of the proof, we show how to find such $H$ in polynomial time using dynamic programming.
	
	To simplify our task, let us first make $G$ connected.
	For that, introduce a new vertex $r$ to $G$, and connect this vertex with each connected component of $G$ by a single edge arbitrarily, even if $G$ consists of a single connected component.
	We thus obtained a tree $T$ that differs from $G$ only in the newly-introduced vertex $r$.
	For each $H \subseteq V(G)$, $G[H]=T[H]$.
	Hence, we are now looking for a subgraph in $T$ on at least $p$ vertices, such that this subgraph does not contain $r$, and we want its total deficiency to be minimum possible.
	
	Make $T$ rooted in $r$.
	For $v_i \in V(T)$, by $T_v$ we denote the subtree of the vertex $v_i$ in $T$.
	Let $t_i$ be the number of child vertices of $v_i$ in $T$.
	By $c_{i,1}, c_{i,2}, \ldots, c_{i,t_i}$ denote the children of $v_i$ in $T$ in arbitrary order.
	If $v_i$ is a leaf vertex, $t_i=0$.
	For each $j \in \{0,1,\ldots, t_i\}$ by $T^j_{v_i}$ denote the subtree of $v_i$ in $T$, but including only subtrees of the first $j$ of its children.
	That is, $$T^j_{v_i}=T_{v_i}[\{v_i\} \sqcup V(T_{c_{i,1}}) \sqcup V(T_{c_{i,2}}) \sqcup \ldots \sqcup V(T_{c_{i,j}})].$$
	Clearly, $T_{v_i}=T^{t_i}_{v_i}$.
	
	Now for each $v_i \in V(T)$, each $j \in \{0,\ldots,t_i\}$ and each $s \in \{0,1,\ldots,|V(T^j_{v_i})|-1\}$, let
	$$OPT^j_0(v_i,s)=\min\left\{
	\defc(T^j_{v_i}[S])
	\left|
	\begin{matrix}
	S \subset V(T^j_{v_i}), \\
	|S| = s, \\
	v_i \notin S \\
	\end{matrix}
	\right.
	\right\}.$$
	In other words, $OPT^j_0(v_i,s)$ equals the minimum deficiency of a subgraph of $T^j_{v_i}$ on exactly $s$ vertices, such that it does not contain $v_i$.
	Also denote $OPT_0(v_i,s)=OPT^{t_i}_0(v_i,s)$.
	
	On the other hand, for each $v_i \in V(T)$, each $j \in \{0,1,\ldots, t_i\}$ each $s \in [|V(T^j_{v_i})|]$ and each $d \in \{0,1,\ldots, k\}$, define
	$$OPT^j_1(v_i,s,d)=\min\left\{
	\defc(T^j_{v_i}[S])
	\left|
	\begin{matrix}
	S \subseteq V(T^j_{v_i}), \\
	|S| = s, \\
	v \in S, \\
	\defc_{T^j_{v_i}[S]}(v)=d \\
	\end{matrix}
	\right.
	\right\}. $$
	Thus, $OPT^j_1(v,s,d)$ equals the minimum deficiency of a subgraph of $T^j_{v_i}$ on $s$ vertices, including $v_i$, such that deficiency of $v_i$ in this subgraph equals $d$.
	For some choice of $v_i$, $j$, $s$, $d$, there may be no corresponding subgraphs.
	In such cases, we put $OPT^j_1(v_i,s,d)=\infty$.
	Also denote $OPT_1(v_i,s,d)=OPT^{t_i}_1(v_i,s,d)$.
	
	Finally, for each $v \in V(T)$ and each $s \in \{0,1,\ldots,|V(T_v)|\}$, let
	$$OPT(v_i,s)=\min\left\{OPT_0(v_i,s),\; \min_{d=0}^k OPT_1(v_i,s,d)\right\}$$
	denote the minimum deficiency of a subgraph of $T_{v_i}$ on $s$ vertices.
	
	Clearly, the minimum possible deficiency we are looking for equals $$\min\limits_{i=p}^{|V(G)|}OPT_0(r,i),$$
	and it is enough to compare this value with $2b$ to solve the initial instance of \textsc{Edge $k$-Core}.
	We now show how we compute the values of $OPT$.
	We do this in a bottom-up manner, starting from the leaf vertices of $T$.
	
	Let $v_i \in V(T)$ be a vertex in $T$.
	Since $T^0_{v_i}$ consists of a single vertex, the only choice of $s$ for $OPT^0_0(v_i,s)$ is $s=0$, and $OPT^0_0(v_i,0)=0$.
	For $OPT^0_1(v_i,s,d)$, the only choice is $s=1$ and $d=k$, and $OPT^0_1(v_i,1,k)=k$, as $\defc(T^0_{v_i})=k$.
	If $v_i$ is a leaf vertex, then computations for $v_i$ are finished, as $OPT_0(v_i,s)=OPT^0_0(v_i,s)$ and $OPT_1(v_i,s,d)=OPT^0_1(v_i,s,d)$.
	
	Otherwise, $t_i > 0$ and all values of $OPT$ for the children of $v_i$ are already computed.
	Consider computing $OPT^j_0(v_i,s)$ for $j>0$.
	Take $S \subset T^j_{v_i}$, $v_i \notin S$.
	Since subtrees of the children of $v_i$ are connected only through $v_i$, it is true that
	\begin{equation}\label{equ:trees:dp}
	\defc(T^j_{v_i}[S])=\defc(T^{j-1}_{v_i}[S\cap V(T^{j-1}_{v_i})])+\defc(T_{c_{i,j}}[S\cap V(T_{c_{i,j}})]).
	\end{equation}
	In some sense, we can minimize total deficiencies of $T^{j-1}_{v_i}$ and $T_{c_{i,j}}$ separately.
	We obtain $$OPT^j_0(v_i,s)=\min_{s_i=0}^{\min\{s,|V(T_{c_{i,j}})|\}} \left( OPT^{j-1}_0(v_i,s-s_i) + OPT(c_{i,j}, s_i)\right).$$
	Thus, $OPT^j_0(v_i,s)$ is computed in $\O(|T_{c_{i,j}}|)$ time.
	We compute these values in increasing order of $j$.
	Since there are $|T^j_{v_i}|$ choices of $s$ for a fixed $j$, computing all values of $OPT_0$ for $v_i$ in total takes $\O(|T^0_{v_i}|\cdot|T_{c_{i,1}}|+|T^1_{v_i}|\cdot|T_{c_{i,2}}|+\ldots+|T^{t_i-1}_{v_i}|\cdot|T_{c_{i,t_i}}|)$ time.
	
	We now turn onto computing $OPT^j_1(v_i,s,d)$ for $j>0$.
	Take $S\subseteq V(T^j_{v_i})$, $v_i \in S$.
	Denote $S_{0}=S\cap V(T^{j-1}_{v_i})$ and $S_1=S\cap V(T_{c_{i,j}})$, so $S=S_0\sqcup S_1$.
	$v_i$ is now included in the subgraph $T^j_{v_i}[S]$, and it may now influence the deficiency of its child $c_{i,j}$.
	Thus, equation \ref{equ:trees:dp} is not quite correct in this case.
	If $u_i \in S$ and $\defc_{T_{c_{i,j}}[S_1]}(c_{i,j})>0$, then in $T^j_{v_i}$ deficiency of $c_{i,j}$ decreases by one because of the edge $v_i c_{i,j}$.
	If $\defc_{T^{j-1}_{v_i}[S_0]}(v_i)>0$, then the deficiency of $v_i$ in $T^j_{v_i}$ also decreases by one.
	Thus, if $c_{i,j} \in S$,
	\begin{multline}\label{equ:trees:dp1}
	\defc(T^j_{v_i}[S])=\defc(T^{j-1}_{v_i}[S_0])-\min(1,\defc_{T^{j-1}_{v_i}[S_0]}(v_i))\\+\defc(T_{c_{i,j}}[S_1])-\min(1,\defc_{T_{c_{i,j}}[S_1]}(c_{i,j})),
	\end{multline}
	and if $u_i \notin S$, equation \ref{equ:trees:dp} holds true.
	Denote by
	$$JOIN(c_{i,j},s_i)=\min\left\{OPT_1(c_{i,j},s_i,0), \min_{d_i=1}^k \left(OPT_1(c_{i,j},s_i,d_i)-1\right)\right\}$$
	the minimum total deficiency of a subgraph of $T_{c_{i,j}}$ on $s_i$ vertices including $c_{i,j}$, but with deficiency of $c_{i,j}$ decreased by one, if it is non-zero.
	This corresponds to the second half of the right part of equation \ref{equ:trees:dp1}.
	Each value $JOIN(c_{i,j}, s_i)$ can be computed in $\O(k)$ time, and there are $|T_{c_{i,j}}|$ choices of $s_i$ for any $j \in [t_i]$.
	Hence, all values of $JOIN$ for the children of $v_i$ together can be computed in total $\O(|T_{c_{i,1}}|\cdot k+|T_{c_{i,2}}|\cdot k+\ldots+|T_{c_{i,t_i}}|\cdot k)=\O(|T_{v_i}|\cdot k)$ running time.
	We then obtain
	\begin{multline}\label{equ:trees:opt1}
	OPT^j_1(v_i,s,d)=\min_{s_i=0}^{\min\{s,|V(T_{c_{i,j}})|\}}\{\\
	OPT^{j-1}_1(v_i,s-s_i,d)+OPT_0(c_{i,j},s_i),\\
	OPT^{j-1}_1(v_i,s-s_i,d+1)+JOIN(c_{i,j}, s_i)\\
	\},
	\end{multline}
	for any $s \in [|V(T^j_{v_i})|]$ and any $d < k$.
	The first argument of $\min$ in equation \ref{equ:trees:opt1} corresponds to $u_i \notin S$ and equation \ref{equ:trees:dp}, and the second argument corresponds to $u_i \in S$ and equation \ref{equ:trees:dp1}.
	If $d=k$, i.e.\ the deficiency of $v_i$ in $S$ equals $k$, then necessarily $u_i \notin S$.
	Thus, in case when $d=k$, we compute $OPT^j_1(v_i,s,d)$ according to equation \ref{equ:trees:opt1}, but without the second argument of $\min$.
	This finishes the description of formulas for computing the values of $OPT_1$.
	Each separate value of $OPT^j_1(v_i,s,d)$ is computed in $\O(|T_{c_{i,j}}|)$ time.
	There are $|T^j_{v_i}|\cdot (k+1)$ choices of $(s, d)$, so computing all values of $OPT^j_1$ for $v_i$ together takes \linebreak $\O(k\cdot \sum_{j=0}^{t_i} |T^j_{v_i}|\cdot |T_{c_{i,j}}|)$ running time.
	Note that this also covers the running time needed for computing the values of $JOIN$ for all children of $v_i$.
	
	It is left to show that the total running time of the algorithm is $\O(k\cdot |V(T)|^2)=\O(k\cdot |V(G)|^2)$.
	Computing the values of $OPT_0$ and $OPT_1$ for a fixed vertex $v_i \in T$ takes $\O(k\cdot \sum_{j=0}^{t_i} |T^j_{v_i}|\cdot |T_{c_{i,j}}|)$ running time.
	We need to show that $$\sum_{v_i \in V(T)}\sum_{j=0}^{t_i} |T^j_{v_i}|\cdot |T_{c_{i,j}}| \le |V(T)|^2,$$
	which is equivalent to
	$$\sum_{v_i \in V(T)}\sum_{j=0}^{t_i} |T^j_{v_i} \times T_{c_{i,j}}| \le |V(T) \times V(T)|.$$
	Note that all sets $T^j_{v_i} \times T_{c_{i,j}}$ are pairwise-disjoint, and each of them is a subset of $V(T) \times V(T)$.
	Hence, the last inequality is essentially true.
	Thus, the dynamic programming itself is done in $\O(k \cdot |V(G)|^2)$ running time by the algorithm.
	Any other part of the algorithm, including constructing and rooting $T$ and finding and comparing $\min_{i=p}^{|V(G)|}OPT_0(r,i)$ with $2b$, takes $\O(|V(G)|)$ time, which is covered by $\O(k\cdot |V(G)|^2)$.
	This finishes the proof.
\end{proof}

For the remaining part of this section we focus on the proof of Theorem~\ref{thm:trees_optimal}.
\fi


\begin{proof}[Proof of Theorem~\ref{thm:trees_optimal}]
	
	The theorem says that completion of $T$ to a graph of total deficiency $0$ can be done using $\lceil \frac{1}{2}\defc(T)\rceil$ edge additions.
	Note that this bound is tight because a single edge addition decreases the total deficiency by at most two.
	When $\defc(T)$ is   even, we have to prove that it is possible to complete $T$ by adding only good edges. 
	When $\defc(T)$ is   odd, 
	we have to complete $T$ to a graph of total deficiency $1$ adding $\lfloor \frac{1}{2} \defc(T) \rfloor$ good 
	edges  and then add one bad edge. Fixing deficiency $1$ with one bad edge is always possible, since the only deficient vertex $u$ has degree $k - 1$ and so must have a non-neighbor.
	In the case $k \ge 4$ this can be also done in a way that connects $u$ to the already added good edges. Thus, from now on, it suffices to prove that we can add $\lfloor \frac{1}{2} \defc(T) \rfloor$ good 
	edges, in a connected way for $k \ge 4$.

	For $k = 1$, vertices with non-zero deficiency are exactly the isolated vertices of $T$. In this case pairing isolated vertices arbitrarily provides the required $\lfloor \frac{1}{2} \defc(T) \rfloor$ good 
	edges.
	
	For $k \ge 2$, it is sufficient to prove the theorem statement for the case when $T$ is connected, i.e.\ $T$ is a tree.
	If $T$ is a forest consisting of at least two trees, one may reduce the number of trees in $T$.
	This can be done by picking two leaf vertices of distinct connected components in $T$ and adding an edge between them.
	Clearly, such an edge addition is good since any leaf vertex has non-zero deficiency, and it reduces the number of connected components in $T$.

	Moreover, for $k = 2$, vertices with non-zero deficiency are exactly the leaves of $T$. Since $T$ is a tree with at least three vertices, an edge connecting any two leaves can be added. Thus, as in the case $k = 1$, pairing the leaves arbitrarily suffices.

	

	
	Now, for every integer  $k \ge 3$, we prove Theorem~\ref{thm:trees_optimal} by induction on the number of vertices in the tree. The fact that the graph on the added edges must be connected in the case $k \ge 4$ will be useful for the induction.
	
	\emph{Base case.}
	Let $T$ be a tree on $n=k+1$ vertices. The only way to complete $T$ to a graph of minimum degree $k$ is to turn it into a complete graph, i.e.\ add every possible missing edge between vertices in $V(T)$.
	Clearly, each edge addition in such completion is good, thus the completion requires exactly $\frac{1}{2}\defc(T)$ edge additions. Suppose there are two connected components formed by the added edges.
	Then $T$ must contain all edges between these components, so it also contains a cycle, since each of the components has at least two vertices. Thus the connectivity condition must be satisfied.

	
	
	\emph{Inductive step.}
	Suppose that Theorem~\ref{thm:trees_optimal} holds for all trees on $n$ vertices, and let $T$ be a tree on $n+1$ vertices.
	We prove that Theorem~\ref{thm:trees_optimal} holds for $T$. Let $v$ be a leaf of $T$ and 
	let $T' =T-v$ be the tree obtained by deleting  $v$ from $T$.
	 By the  induction hypothesis, $T'$ can be completed to a graph of total deficiency $(\defc(T')\bmod 2)$ using $\lfloor \frac{1}{2}\defc(T') \rfloor$ edge additions.
	Let $A'$ be the graph on the deficient vertices of $T'$ formed by the good edges added during the completion.

    Our ultimate goal is to transform $A'$ in such a way that it accounts for the new vertex $v$ as well. We shall do this by first removing edges from $A'$, and then adding good edges between vertices which are not yet adjacent.
    In the case $k \ge 4$, we must also end up with a connected graph on the added edges.

	Briefly explained, our technique of adding and removing edges is as follows.
	Take an edge $st \in E(A')$, such that 1) $s \neq v$, $t \neq v$ and 2) $sv$ and $tv$ are not yet in the graph.
	Delete the edge $st$, and add both edges $sv$ and $tv$.
	This operation preserves deficiencies of both $s$ and $t$, while it decreases the deficiency of $v$ by two.
	Note that $s$ and $t$ also remain connected through $v$.
	We can do the same with a matching instead of a single edge, thus we need a matching of size roughly $k / 2$ to nullify the deficiency of $v$.

	The rest of the proof is structured in two parts. First, we show that there is indeed a sufficiently large matching in $A'$.
	Second, we give a detailed description of how to reroute the edges of the matching to the new vertex $v$, and carefully verify the correctness of the procedure.

	\textbf{Finding a matching.} We will need the following properties of $A'$.
	
	\begin{equation}
		 \text{If $k \ge 4$, $A'$ is connected.}\label{enum:trees:connected}
	\end{equation}	
	The correctness of 	\eqref{enum:trees:connected}  follows from the induction hypothesis.
		Because each vertex in $T'$ has deficiency at most $k-1$ and each edge addition is good, we have that 
\begin{equation}
	\label{enum:trees:maxdeg} \emph{$\Delta(A')\le k-1$.}
	\end{equation}	
		Also 			
			\begin{equation}
		\label{enum:trees:manyedges} \emph{$|E(A')|\ge \frac{n(k-2)+1}{2}$,}
		\end{equation}	
		since there must be at least $\frac{nk - 1}{2}$ edges in the graph after the completion to deficiency $(\defc(T')\bmod 2)$, and only $n - 1$ of the edges are in $T'$.
		
		
	\begin{equation}
		\emph{$|V(A')|\ge k$. If $n>k+1$, then $|V(A')|>k$.}\label{enum:trees:manyvertices}
	\end{equation}
	\ifshort
\else
    To prove~\eqref{enum:trees:manyvertices}, 		
		suppose that $|V(A')|\le k-1$.
		By   \eqref{enum:trees:maxdeg}, $2\cdot |E(A)|\le \Delta(A') \cdot |V(A')| \le (k-1)^2$.
		Then by  \eqref{enum:trees:manyedges},  we have that either $(k-1)^2\ge n(k-2)+1$, or $k^2-2k\ge n(k-2)$. Hence   $k \ge n$, which is a contradiction. Therefore,  $|V(A')|\ge k$.
		
		Suppose now that $|V(A')|=k$, but $n\ge k+2$.
		Then either $k(k-1)\ge n(k-2)+1 \ge (k+2)(k-2)+1$, or $k^2-k \ge k^2-3$, so $k\le 3$.
		Thus, $k=3$ and $n=k+2=5$ (if $n>k+2$, $k$ should be strictly less than $3$).
		Since $|V(A')|=k=n-2$, $T$ should have two vertices of degree at least three.
		But then $T$ should contain at least five edges, but it only contains four.
		A contradiction.
    \fi
	
	We now use these properties of $A'$ to show that there is a large matching in $A'$.
	For lower bounds on the size of a maximum matching we rely on the recent work of Henning and Yeo~\cite{Henning2018}.
	
	\begin{proposition}[\cite{Henning2018}]\label{thm:matching_lbs}
		For any integer $t \ge 3$, any connected graph $G$ with $|V(G)|=n$, $|E(G)|=m$ and $\Delta(G) \le t$, contains a matching of size at least
		$$\left(\frac{t-1}{t(t^2-3)}\right)n+\left(\frac{t^2-t-2}{t(t^2-3)}\right)m-\frac{t-1}{t(t^2-3)}, \text{ if }t\text{ is odd,}$$
		or at least
		$$\frac{n}{t(t+1)}+\frac{m}{t+1}-\frac{1}{t}, \text{ if }t\text{ is even.}$$
	\end{proposition}
	
	We shall use Proposition \ref{thm:matching_lbs} to show that $A'$ contains a matching of size roughly $\frac{k}{2}$,
	as stated in the following claim.
	
	\begin{claimO}\label{claim:trees:largemt}
	When $k$ is odd and $n=k+1$, $A'$ has a matching of size at least $\frac{k-1}{2}$. Otherwise, 
		$A'$ has a matching of size at least $\lceil \frac{k}{2} \rceil$.
	\end{claimO}

    \ifshort
\else
	\begin{proof}
		Consider the case when $k$ is even.
		We apply Proposition \ref{thm:matching_lbs} with  $t=k-1$ and $A'$, and use 
		properties \eqref{enum:trees:connected} and \eqref{enum:trees:maxdeg}.
		Since the lower bound from the theorem statement is rounded up, we need to show that
		$$\left(\frac{t-1}{t(t^2-3)}\right)|V(A')|+\left(\frac{t^2-t-2}{t(t^2-3)}\right)|E(A')|-\frac{t-1}{t(t^2-3)}>\frac{k}{2}-1.$$
		
		By properties \eqref{enum:trees:manyedges} and \eqref{enum:trees:manyvertices}, and by replacing $k$ with $t+1$, it is sufficient  to show that
		\begin{equation}\frac{(t-1)(t+1)}{t(t^2-3)}+\left(\frac{t^2-t-2}{t(t^2-3)}\right)\cdot\left(\frac{n(t-1)+1}{2}\right)-\frac{t-1}{t(t^2-3)}>\frac{t+1}{2}-1, \label{eq:lemma_matching}\end{equation}
		for any odd $t \ge 3$ and any $n \ge k+1=t+2$.
		After multiplying both parts of  \eqref{eq:lemma_matching} by $2t(t^2-3)$, we obtain
		$$2(t-1)(t+1)+(t^2-t-2)\cdot\left(n(t-1)+1\right)-2(t-1)>\left({t+1}-2\right)\cdot t(t^2-3),$$
		which is simplified to
		$$n \cdot (t^3-2t^2-t+2)>t^4-t^3-6t^2+6t+2.$$
		Since $n \ge t+2$, it is  sufficient to show that
		$$(t+2) \cdot (t^3-2t^2-t+2)>t^4-t^3-6t^2+6t+2,$$
		or that 
		$$t^3+t^2-6t+2>0.$$
		Since $t>0$, we weaken the last inequality to $$t^3-6t>0,$$ which holds true for $t>\sqrt{6}$.
		Thus, for $t\ge 3$, inequality \eqref{eq:lemma_matching} holds.
		We hereby have shown that when $k$ is even, $A'$ has a matching of size $\frac{k}{2}$.
		
		Consider now the case of $k=3$.
		We need to consider this case separately because $t=k-1=2$ does not follow from  Proposition \ref{thm:matching_lbs}.
		When $k=3$, we need a matching of size two in $A'$.
		By  \eqref{enum:trees:connected} and \eqref{enum:trees:maxdeg}, $A'$ is a connected graph of maximum degree two.
		Hence, $A'$ is either a simple cycle or a simple path.
		By property \ref{enum:trees:manyedges}, $A'$ is a cycle or a path consisting of at least $\lceil \frac{(k+1)(k-2)+1}{2}\rceil=\lceil \frac{4\cdot 1+1}{2}\rceil=3$ edges.
		Clearly, if $A'$ is a path on at least three edges or a cycle on at least four edges, $A'$ has a matching of size two.
		The only option left for $A'$ is to be a cycle on three vertices and edges, so $|V(A')|=3=k$.
		By property \ref{enum:trees:manyvertices}, this is only possible if $n=k+1$.
		Thus, if $n>k+1$, $A'$ has a matching of size at least two.
		If $n=k+1$, $A'$ is only guaranteed to have a matching of size $1=\frac{k-1}{2}$.

		It is left to consider the case of odd $k \ge 5$.
		Again, apply Proposition \ref{thm:matching_lbs} to $A'$ with even $t=k-1$. We need to show that $A'$ has a matching of size at least $\frac{k+1}{2}=\frac{t+2}{2}$, so it is sufficient to show that
		$$\frac{|V(A')|}{t(t+1)}+\frac{|E(A')|}{t+1}-\frac{1}{t}>\frac{t+2}{2}-1.$$
		According to   \eqref{enum:trees:manyedges} and \eqref{enum:trees:manyvertices}, it is enough to show that
		$$\frac{t+1}{t(t+1)}+\frac{1}{t+1}\cdot\frac{1}{2}\cdot(n(t-1)+1)-\frac{1}{t}>\frac{t+2}{2}-1,$$
		$$\frac{(t-1)n+1}{2t+2}>\frac{t}{2}.$$
		Multiplying both parts by $2t+2$ and using $n\ge k+2=t+3$, we obtain
		$$(t-1)(t+3)+1>t(t+1),$$
		or, equivalently, 
		$t>2.$
		Thus, when $k \ge 5$ is odd and $n\ge k+2$, $A'$ has a matching of size $\frac{k+1}{2}$.
		
		When $n=k+1$, we need a matching of size $\frac{k-1}{2}=\frac{t}{2}$, so we need to show that 
		$$\frac{|V(A')|}{t(t+1)}+\frac{|E(A')|}{t+1}-\frac{1}{t}>\frac{t}{2}-1,$$
		or
		$$\frac{(t-1)n+1}{2t+2}>\frac{t}{2}-1,$$
		This is equivalent to $$t^2+t-1>t^2-t-2,$$
		which is true for $t\ge0$.
		This completes the  proof of the lemma.
	\end{proof}
    \fi
	
	\textbf{Rerouting the edges.} Now we shall use the matching provided by Claim~\ref{claim:trees:largemt} to conclude the inductive step.
	Denote by $G'$ the graph obtained after the completion of $T'$ to a graph of total deficiency $(\defc(T')\bmod 2)$.
	That is, $V(G')=V(T')$ and $E(G')=E(T')\sqcup E(A')$.
	If $\defc(T')$ is odd, $G'$ has a single vertex with deficiency one, denote it by $u$.
	For every other vertex $s \in V(G')$, $\defc_{G'}(s)=0$.
	Our goal is to transform $G'$ into a graph $G$ that will correspond to the graph obtained after the completion of $T$ using only good edge additions.
	
	We initialize $G$ with $G'$.  Let us remind that $v$ is a leaf of $T$ and  $T'=T-v$. We  
	denote the only neighbor of $v$ in $T$ by $p$. Since 
	$G$ is missing vertex $v$, we  introduce $v$ to $G$, which is now isolated in $G$.
	Now $V(G)=V(T)$, so it is left to add missing edges to $G$, while probably removing some of the existing edges.
	Of course, these added edges should include the edge $pv$, since $E(T)\subseteq E(G)$ must hold.
	Similarly, we should not remove any edges of $T'$ from $G$.
	Thus, we can remove edges in $E(A')$ only. We denote by $A$ the graph of added edges in $G$, analogously to $A'$ in $G'$.
	
	As was explained before, our basic technique is to remove the edges of the matching in $A'$, and connect their endpoints to $v$. However, there are several issues to deal with.
	First, if $p$ is in $V(A')$, we have to ensure that one of the edges in $E(A')$ incident to $p$ gets removed, otherwise one of the edge additions is wasted on $p$. This edge removal may in turn disconnect $A'$.
	Second, depending on the parity of $\defc(T')$ we may have to deal with the already-deficient vertex $u$ of $G'$, and the parity of $k$ comes into play as well. 
	Thus, in the rest of the proof we go over five different cases and show that in each of them the rerouting is possible.
	\ifshort
\else
	We start with the cases where $k$ is even.
	Recall that we always start with $G$ being a copy of $G'$ with isolated vertex $v$.
	\fi

	
	\begin{figure}[!ht]
		\centering
		\begin{tabular}{ccc}
			(c)
			\begin{tabular}{l}
				\begin{tikzpicture}[every node/.style={draw=black,fill,thick,circle,inner sep=0pt,minimum size=0.2cm}]
				\node (a) at (0,0) {};
				\node[label=left:$p$] (b) at (0.6, -0.7) {};
				\node[label=left:$u$] (c) at (-0.3, -0.8) {};
				\node (d) at (0.2, -1.6) {};
				\node (e) at (0.9, -1.8) {};
				\node (f) at (1.4, -1.4) {};
				\node[label=right:$v$] (g) at (0.7, -3) {};
				\draw (a) -- (c);
				\draw (a) -- (b);
				\draw (b) -- (d);
				\draw (b) -- (e);
				\draw (b) -- (f);
				\draw[red] (a) to[bend left] (f);
				\draw[red,thick] (f) to[bend left] (e);
				\draw[red,thick] (e) to[bend left] (d);
				\draw[red,ultra thick] (d) to[bend left] (c);
				\draw[dashed] (g) -- (b);
				\draw[dashed,red,thick] (g) to[bend left] (c);
				\draw[dashed,red,thick] (g) to[bend left] (d);
				\end{tikzpicture}
			\end{tabular}
			&
			(d)
			\begin{tabular}{l}
				\begin{tikzpicture}[every node/.style={draw=black,fill,thick,circle,inner sep=0pt,minimum size=0.2cm}]
				\node (b) at (0.6, 0) {};
				\node (c) at (-0.3, -0.8) {};
				\node (d) at (0.2, -1.6) {};
				\node (e) at (0.9, -1.8) {};
				\node[label=right:$p$] (f) at (1.4, -1.4) {};
				\node[label=right:$v$] (g) at (1.5, -2.8) {};
				\draw (b) -- (c);
				\draw (b) -- (d);
				\draw (b) -- (e);
				\draw (b) -- (f);
				\draw[red,ultra thick] (f) to[bend left] (e);
				\draw[red,thick] (e) to[bend left] (d);
				\draw[red, thick] (c) to[in=90, out=45,bend left] (f);
				\draw[red, thick] (d) to[bend left] (c);
				\draw[dashed] (g) -- (f);
				\draw[dashed,red,thick] (g) to[bend left] (e);
				\end{tikzpicture}
			\end{tabular}
			&
			(e)
			\begin{tabular}{l}
				\begin{tikzpicture}[every node/.style={draw=black,fill,thick,circle,inner sep=0pt,minimum size=0.2cm}]
				
				\node (a) at (0,0) {};
				\node (b) at (0.6, -0.7) {};
				\node[label=left:$u$] (c) at (-0.4, -1.4) {};
				\node[label=right:$p$] (d) at (0.2, -1.6) {};
				\node (e) at (0.9, -1.8) {};
				\node (f) at (1.6, -1.4) {};
				\node[label=right:$v$] (g) at (-0.1, -3) {};
				\draw (a) -- (c);
				\draw (a) -- (b);
				\draw (b) -- (d);
				\draw (b) -- (e);
				\draw (b) -- (f);
				\draw[red,ultra thick] (a) to[bend left] (f);
				\draw[red,thick] (f) to[bend left] (e);
				\draw[red,thick] (e) to[bend left] (d);
				\draw[red,ultra thick] (d) to[bend left] (c);
				\draw[dashed] (d) -- (g);
				\draw[dashed,red,thick] (c) to[bend right] (g);
				\draw[dashed,red,thick] (g) to[out=180,in=180] (a);
				\draw[dashed,red,thick] (c) to[bend left] (f);
				\end{tikzpicture}
			\end{tabular}
		\end{tabular}
		\caption {The cases of rerouting for $k=3$. Solid edges denote the edges of $G'$. Straight black edges denote the edges of $T'$, and curved red edges denote the edges of $A'$. Edges of the matching in $A'$ that are deleted in $G$ are highlighted bold. Dashed edges denote the newly-added edges in $G$.}
		\label{fig:trees:oddk}
	\end{figure}
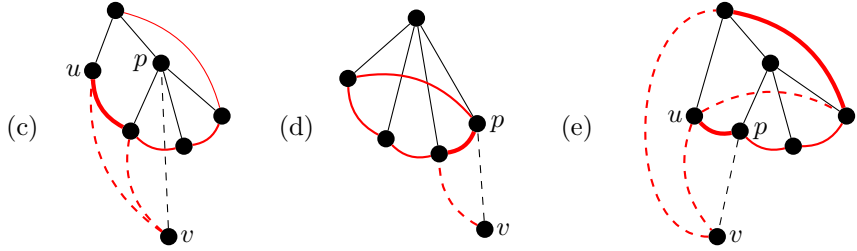

    \ifshort
\else

	\textbf{Case (a).}
	\emph{$k$ is even and $p \in V(A')$.}
	By Claim~\ref{claim:trees:largemt}, there is a matching of size $\frac{k}{2}$ in $A'$.
	Denote this matching by $M$.
	Since $p \in V(A')$, we may ensure that $p \in M$.
	Suppose that $p \notin M$, $p$ is incident to at least one edge in $E(A')$, say $pq$.
	If $q \in M$, remove edge covering $q$ from $M$.
	If $q \notin M$, remove an arbitrary edge from $M$.
	Finally, add edge $pq$ to $M$.
	Now $M$ is a matching of size $\frac{k}{2}$ and $p \in M$.
	We want to remove all edges of $M$ from $G$, and add all $2\cdot \frac{k}{2}=k$ edges connecting $v$ with the vertices covered by $M$.
	Then the degree of $v$ in $G$ is exactly $k$, so $\defc_G(v)=0$, no deficiency of any other vertex has changed, and $G$ now contains the edge $pv$.
	However, the graph formed by the good edges $A$ must be connected, and this may not be the case since we replace the edge $pq$ by edges $pv$ and $qv$, and $pv$ belongs to $T$ so $p$ and $q$ are not necessarily connected through the edges of $A$.

    So before replacing the edges of $M$, we tweak it in order to preserve the connectivity. If $p$ has degree one in $A'$ then after replacing the edges $p$ is not covered by the good edges anymore. If $p$ has a neighbor $t$ in $A'$ which has degree one in $A'$, then we take the edge $pt$ instead of $pq$ in $M$. Since $v$ is going to have at least three incident edges, there will be at least one edge of $A$ among them which is not $vt$, and thus $t$ will be connected to the rest of $A$. If removing $pq$ makes $A$ disconnected and none of the above is the case, the connected component of $p$ in $A' - pq$ has an edge which is not incident to $p$. If one of these edges is already in $M$, then the connectivity will hold since $q$ will be adjacent to $v$ and a vertex in the same component as $p$ will be adjacent to $v$.
    Otherwise we remove from $M$ any edge which is not $pq$ and add any edge which belongs to the connected component of $p$ in $A' - pq$ and is not incident to $p$.

	Finally, after rerouting the edges of $M$ the obtained graph $G$ corresponds to an appropriate completion of $T$.
	Note that if $G'$ contains the vertex $u$ with deficiency one, it remains deficient in $G$ as well.

	\textbf{Case (b).}	
	\emph{$k$ is even and $p \notin V(A')$.}
	The difference with the previous case is that now we cannot simply ensure that edge $pv$ gets added to $G$.
	Add edge $pv$ to $G$.
	This does not change the deficiency of $p$, since it is a vertex of zero deficiency, but deficiency of $v$ now equals $k-1$.
	Take a matching $M$ of size $\frac{k}{2}$ in $A'$, according to Claim~\ref{claim:trees:largemt}.
	If $G$ contains the unique vertex $u$ of deficiency one, and $u \in M$, remove the edge covering $u$ from $M$.
	Otherwise, remove an arbitrary edge from $M$.
	$M$ is now a matching of size $\frac{k}{2}-1$ with $u \notin M$, if $u$ exists.
	Note that $p \notin M$ necessarily, as $p \notin V(A')$.
	Remove all edges of $M$ from $G$, and add all $k-2$ edges connecting $v$ with vertices covered by $M$.
	$v$ is now a vertex of deficiency one in $G$.
	If $u$ exists, connect $u$ and $v$ by a new edge.
	Clearly, this is a good edge addition.
	This edge have not existed before, because we made sure that $u \notin M$.
	If $u$ does not exist, $v$ is a single vertex of deficiency one in $G$.
	In any case, $G$ is an appropriate completion of $T$.
	
	We have hereby proved Theorem~\ref{thm:trees_optimal} for the case of even $k$.
	We now turn to the cases where $k$ is odd.
	Clarifying pictures for all three cases to consider are presented in Figure \ref{fig:trees:oddk}.
	We start with the easier case.
	
	\textbf{Case (c).}
	\emph{$k$ is odd and $p \notin V(A')$}.
	As in Case (b), introduce the edge $pv$ to $G$, so $\defc_G(v)=k-1$, and the deficiency of $p$ does not change.
	By Claim~\ref{claim:trees:largemt}, there is a matching $M$ of size $\frac{k-1}{2}$ in $A'$.
	As usual, remove all edges of $M$ from $G$, and add all $k-1$ edges connecting $v$ with the endpoints of edges in $M$.
	The vertex $v$ has zero deficiency now, and for any other vertex the deficiency is the same.
	Clearly, $G$ is an appropriate graph.
	
	\textbf{Case (d).}
	\emph{$k$ is odd and $p \in V(A')$, there is no deficient vertex in $G'$.}
	Take a matching $M$ of size $\frac{k-1}{2}$ in $A'$, and ensure that $p \in M$. In the case $k \ge 5$, do the same tweaking to $M$ as in Case (a) to ensure connectivity.
	Remove all edges of $M$ from $G$ and connect $v$ to all $k-1$ vertices covered by $M$.
	The edge $pv$ is now contained in $G$, and $v$ is the only vertex with deficiency one in $G$.
	This is an appropriate completion of $T$.
	
	\textbf{Case (e).}
	\emph{$k$ is odd and $p \in V(A')$, $\defc_{G'}(u)=1$.}
	Note that $n \ge k+2$ in this case, since for $n=k+1$ there may be no deficient vertices in $G'$.
	Take a matching $M$ of size $\frac{k+1}{2}$ in $A'$ according to Claim~\ref{claim:trees:largemt}, and ensure that $p \in M$. We consider two cases, either $up \in M$ or not.
	In the case $up \in M$, if there is another neighbor $q$ of $p$ in $A'$ which is not covered by $M$, replace $up$ in $M$ by $qp$, and thus we reduce to the second case. Otherwise, remove all edges of $M$ from $G$.
	Now $\deg_G(u)=k-2$, so at least two vertices covered by $M$ are not adjacent to $u$.
	At least one of these vertices is distinct from $p$, denote this vertex by $w$.
	Add edges connecting $v$ to each vertex covered by $M$, except for the vertex $w$.
	Thus, $v$ gets connected to exactly $\frac{k+1}{2}\cdot 2-1=k$ vertices, including $p$ and $u$, so $\defc_G(v)=0$.
	Finally, add an edge connecting $u$ and $w$.
	Now each vertex in $G$ is of zero deficiency.
	Clearly, each edge addition in this construction is good. To see that $A$ is connected in the case $k \ge 5$, first note that only removing edges $up$ and $wt$ could influence connectivity, where $t$ is such that $wt \in M$. For all other edges of $M$ which were removed, their endpoints are adjacent to $v$ in $A$, and thus connected in $A$. The pair $w$ and $t$ remains connected in $A$ since $t$ is adjacent to $v$, and $w$ is connected to $v$ through $u$. The vertex $p$ either had the only neighbor $u$ in $A'$, then $p$ is not in $V(A)$, or there was another neighbor $q$ of $p$ in $A'$. Note that $q$ is necessarily covered by $M$ since we could not replace the edge $up$. Then $q$ is adjacent to $v$ in $A$, and $p$ and $u$ are connected through $q$ and $v$.
	Also $pv \in E(G)$, so $G$ is an appropriate completion of $T$.

	If $up \notin M$, first remove from $M$ the edge covering $u$, or if there is no such edge, remove an arbitrary edge from $M$ so that $p$ remains covered by $M$.
	Remove all edges of $M$ from $G$ and connect $v$ to all endpoints of these edges as usual.
	Now $\defc_G(u)=\defc_G(v)=1$, and $uv \notin G$, so add a good edge between $u$ and $v$. After that, $G$ is an appropriate completion of $T$ with no deficient vertices. To see that $A$ is connected, we only need to check that $p$ and its pair $q$ in $M$ remain connected. Either $u$ is in the connected component of $p$ in $A$, and then $p$ and $q$ are connected through $u$ and $v$, or we do the same tweaking to $M$ as in Case (a) before actually removing the edges.
	
	We considered each case required to accomplish  the inductive step.
	This concludes the proof of Theorem~\ref{thm:trees_optimal}.
	\fi
\end{proof}



Since the class of forests is exactly the class of $1$-degenerate graphs, it is reasonable to ask whether \textsc{Edge $k$-Core} is polynomially solvable on other classes of graphs of bounded degeneracy.
The answer is negative, and it was shown by Chitnis and Talmon in \cite{Chitnis2018}, where they provided a reduction from \textsc{Clique} to \textsc{Edge $k$-Core}.
We note that they used this reduction to prove that \textsc{Edge $k$-Core} is $\W[1]$-hard when parameterized by the combined parameter $b+p$, even when $k=3$.

\begin{proposition}[\cite{Chitnis2018}]
	\textsc{Edge $k$-Core} is \NP-hard even on the class of $2$-degenerate graphs for $k=3$.
\end{proposition}

\section{Vertex Cover}\label{sec:VC}




This section is dedicated to \probCore parameterized by the minimum size of a vertex cover of the input graph $G$.
We show that this problem admits an \FPT~algorithm and complement this result by ruling out the existence of a polynomial kernel. 
 We start with the high level description of the main ideas  behind our algorithm.

 \medskip\noindent\textbf{High-level description of the algorithm.} In order to prove that \probCore is \FPT \, parameterized 
 by the vertex cover number of the input graph, we construct an \FPT-time Turing reduction from
\probCore to an instance of integer linear program (ILP)  whose number of variables is bounded by some function of the vertex cover. While reducing to \probILPshort  is a common approach in the design of parameterized algorithms, see \cite[Chapter~6]{cygan2015parameterized}, the reduction for \probCore is not straightforward. 
In order to make the whole approach applicable, we need a new combinatorial result, Lemma~\ref{lemma:check_eg}. 
The proof of this lemma  strongly exploits the refinement of Tripathi and Vijay \cite{Tripathi2003} of the classical theorem of Erd\H{o}s and Gallai about degree sequences  \cite{erdHos1960grafok}. 
%
%
%

The reduction target  is the following \probILP
(\probILPshort) problem.

\defparproblema{\probILPshort}%
{Matrix $A \in \mathbb{Z}^{m\times \ell}$ and vector $b \in \mathbb{Z}^{m}$.}%
{$\ell$}
{Is there a vector $ {x} \in \mathbb{Z}^{\ell}$  such that  $A \cdot {x} \le b$?
	}



\probILPshort  is \FPT\, by 
  the celebrated  result of Lenstra \cite{Lenstra1983}.

\begin{proposition}[\cite{Kannan1987,Lenstra1983,Frank1987}]\label{thm:ilpalgo}
	\probILPshort can be solved using $\O(\ell^{2.5\ell+o(\ell)} \cdot L)$ arithmetic operations and space polynomial in $L$.
	Here $L$ is the number of bits in the input.
\end{proposition}

 Let $G$ be a simple undirected graph on $n$ vertices and $b$, $k$, and $p$ be  integers. 
Let $\vc$ be the minimum size of a vertex cover in an $n$-vertex graph $G$. 
Our \FPT \, Turing reduction constructs in time $2^{\Oh(\vc^2)}\cdot \polyn$ at most $2^{\O(\vc^2)}$ instances of 
\probILPshort. Each instance of \probILPshort has $\ell=2^{\Oh(\vc)}$ variables. Moreover, at least one of the constructed instances of  \probILPshort is  a yes-instance   if and only if 
  one can 
build a  $k$-core of size  $p$ in $G$ by adding at most 
 $b$ edges.
Thus by applying Proposition~\ref{thm:ilpalgo} to each of the instances of \probILPshort, we obtain an \FPT\, (parameterized by $\vc$)  algorithm for 
\probCore.



\medskip

Recall that in \probCore we are looking for a vertex subset $H\subseteq V(G)$ of size at least $p$ such that $G[H]$ can be completed to a graph of minimum degree at least $k$ using at most $b$ edge additions. In what follows, we describe the reduction from \probCore to \probILPshort.

We start with computing  a minimum vertex cover $C$ of $G$. It is well-known that a simple branching algorithm does this job in time  $2^{|C|}\cdot\polyn$, see e.g.~\cite{cygan2015parameterized}.
We simplify our task a bit by assuming that $C \subseteq H$: we just branch into  $2^{|C|}$ possible options of $H\cap C$. For each option we  delete   vertices $C\setminus H$ from $G$. We use the following notion of vertex types.
\begin{definition}[Vertex types]
	Let $G$ be a graph and $C$ be its vertex cover.
	For $S \subseteq C$ and a vertex $v\not \in C$, we say that \emph{$v$ has type $S$} if $N_G(v)=S$.
\end{definition}

 We encode the choice of $H$ (up to isomorphism of $G[H]$) using only $2^{|C|}$ positive integers: for each $S \subseteq C$ we just need to indicate how many vertices of type $S$ are in $H$.
That is, the values of $2^{|C|}$ variables $x_S:=|\{v \in H\mid N_G(v)=S, v \notin C\}|$ uniquely define the graph $G[H]$.
Then inequality $|C|+\sum_{S\subseteq C}x_S\ge p$   ensures that $|H|\ge p$.

The non-trivial part of the proof is to encode in \probILPshort  that  
  $G[H]$ can be completed to a $k$-core graph using at most $b$ edges.
In graph  $G[H]$, the vertex set $C$ is a vertex cover and the set $I=H\setminus C$ is an independent set. 
 Assume that $G[H]$ can be completed into a $k$-core graph by making use of  a set of edges $B$, $|B|\leq b$.
The set $B$ can be partitioned into $B=B_C \cup B_I$. Here $B_C$ are the edges with at least one endpoint in $C$, 
and $B_I\subseteq \binom{I}{2}$ are the remaining edges. Every edge of $B_I$ has two endpoints in $I$.
We encode the sets $B_C$ and $B_I$ in \probILPshort in different ways.


It is convenient to assume that $B_C$  contains no  edges with both endpoints in $C$. We reach this condition by branching into 
 $2^{\binom{|C|}{2}}=2^{\Oh(\vc^2)}$ possible options of which edges between vertices in $C$ are added to $G$. For each such guess we also update the value $b$ and the conditions on degrees of vertices in $C$.
 
%
%
%

The next step in the reduction to  \probILPshort  is to encode the graph $G[H] +B_C$. Since we do not have edges with both endpoints in $C$ anymore, $B_C$ consists only of edges between $C$ and $I$.
Since  $C$  is also a vertex cover of $G[H] +B_C$, there are at most $2^{|C|}$ different types of vertices in $H\setminus C$ in the graph  $G[H] +B_C$.
%
 A vertex $v$ of type $S'$ in $G[H]+ B_C$ has type $S \subseteq S'$ in the graph $G[H]$.
Let $y_{S,S'}$ (for $S\subseteq S'\subseteq C$) denote the number of vertices of type $S$ in $G[H]$ that become vertices of type $S'$ in $G[H]+ B_C$.
Then the set of equations $\sum_{S'\supseteq S} y_{S,S'}=x_S$, for each $S\subseteq C$,
ensures that these values correspond to the actual structure of $G[H]$.
The cardinality of $B_C$ is then   encoded as $\sum_{S'\subseteq C}\sum_{S\subseteq S'}|S'\setminus S|\cdot y_{S,S'}.$
Since for each vertex $v \in C$ the graph $G[H]+B_C$ contains all edges incident to $v$ in $G[H]+ B$, the resulting degree of $v$ can be checked immediately.
Formally, $\deg_{G[C]}(v) + \sum_{S'\ni v}\sum_{S\subseteq S'}y_{S,S'}\ge k$ is equivalent to $\deg_{G[H]+ B}(v)\ge k$.

 We proceed with the description of how we encode the edge set $B_I$. 
For  that we need to ensure that for each vertex of $I$ its degree in $G[H]+ (B_C\cup B_I)$ is at least $k$.
Since adding edges between vertices in $I$ could significantly increase the vertex cover of $G[H]$, we cannot do the encoding in the same way as for the edges in $B_C$.
However, $I$ remains to be an independent set in $G[H]+ B_C$. Therefore, $B_I$ can be any set of edges subject to the condition that  in $G[I]+ B_I$ the degree of every vertex $v \in I$ is at least $\defc_{G[H]+ B_C}(v)$.
Thus, to ensure that $B_I$ is an appropriate set all we need to consider are the deficiencies of vertices in $I$.

The deficiencies of vertices in $I$ are integers within the range
$[\max\{0,k-|C|\}, k]$.
Since $G[I]$ is an empty graph,  it is not necessary to   know the deficiency  of each particular vertex in $I$.
Knowing the number of vertices in $I$ of each particular deficiency is sufficient for our purposes.
For $i \in [\max\{0,k-|C|\},k]$, let $s_i$ denote the number of vertices in $I$ with deficiency $i$.
These variables can be encoded with \probILPshort  equations using the variables $y_{S,S'}$.

\medskip
We arrive to  the most interesting and non-trivial part of the reduction. While the inequalities we have built so far  are necessary for encoding  the information about the set  $B_I$,  they are not sufficient. The reason is that not every sequence of  integers corresponds to a sequence of vertex degrees in a graph. There is a classical theorem of Erd\H{o}s and Gallai  providing a characterization of graphic sequences. However, if we use  this theorem to encode  graphic sequences in \probILPshort, the resulting integer program could have unbounded (by a function of $\vc$) number of variables.  To overcome this obstacle, we need Lemma~\ref{lemma:check_eg}, a new combinatorial result about graphic sequences.

We want to encode the property that there exists a set of edges $B_I$ of size at most $b-|B_C|$ such that the edges of $B_I$ form a graph with at least $s_k$ vertices of degree at least $k$, at least $s_{k-1}$ \emph{other} vertices of degree at least $k-1$, and so on down to $s_{\max\{0,k-|C|\}}$. 
One technical obstacle here is that we ask for $s_i$ vertices of degree \emph{at least} $i$, not of degree \emph{exactly} $i$.
In what follows, for clarity,  we explain only how  to encode the existence of an edge set forming a graph with $t_{i}$ vertices of degree exactly $i$ for each $i \in [\max\{0,k-|C|\},k]$. For the ``at least'' case we need to do more work, but the main idea remains the same. 
Note that the case we explain here (requiring $t_i$ vertices of degree exactly $i$) is achieved automatically if all edges in $B_I$ are good edges (that is, consecutive addition of edges from $B_I$ decreases deficiencies of exactly two vertices by one) and the cardinality of this set is found easily as $\frac{1}{2}\sum_{i}t_{i}$.

Let us remind the following classical graph-theoretical notion.

\begin{definition}[Graphic sequences]
A sequence  $d_1, d_2, \ldots, d_n$  of $n$ non-negative integers.
	 is called \emph{graphic} if there exists a graph $G$ with $V(G)=\{v_1, v_2, \ldots, v_n\}$, such that $\deg_G(v_i)=d_i$ for each $i \in [n]$.
\end{definition}

In terms of this notion, our task is to check that a sequence consisting of integers from $ [\max\{0,k-|C|\},k]$, where the integer $i$ appears exactly $t_{i}$ times, is a graphic sequence.
The problem of determining that a given sequence is graphic was approached by Erd\H{o}s and Gallai in their famous work \cite{erdHos1960grafok}.

\begin{proposition}[Erd\H{o}s-Gallai Theorem, \cite{erdHos1960grafok}]\label{thm:erdos-gallai}
	A sequence of non-negative integers $d_1 \ge d_2 \ge \ldots \ge d_n$ is graphic if and only if $\sum_{i=1}^n d_i$ is even and
	$\sum_{i=1}^t d_i \le t \cdot (t-1) + \sum_{j=t+1}^n \min\{d_j, t\}$
  for each $t \in [n]$.
\end{proposition}

However, the statement of Proposition~\ref{thm:erdos-gallai} does not allow us to encode corresponding inequalities in \probILPshort  with the number of variables bounded by $|C|$.  We need a refined version of this proposition, Lemma~\ref{lemma:check_eg}. This 
 combinatorial result on graphic sequences   of integers in a short range is crucial in constructing \probILPshort inequalities with bounded number of variables.
 The proof of the lemma is based on the modification of  the Erd\H{o}s-Gallai theorem due to Tripathi and Vijay~\cite{Tripathi2003}.
%
\begin{lemmaO}\label{lemma:check_eg}
	Let $d_1 \ge d_2 \ge \ldots \ge d_n$ be a sequence of non-negative integers, such that for each $j \in [n]$ $d_j \in [k-a, k]$, for some integers $0 \le a \le k < n$.
	For each $i \in [k-a,k]$, let $t_{i}=|\{j \mid d_j=i\}|$ be the number of integers equal to $i$ in the sequence.
	For each $D \in [k-a,k]$, let $T_D=\sum\limits_{i=D}^k t_{i}$.
	
	Then $d_1, d_2, \ldots, d_n$ is graphic if and only if $\sum_{i=k - a}^k i \cdot t_{i}$ is even and for each $D \in [k-a, k]$ at least one of the following holds:
	\begin{enumerate}
		\item $T_D < k-a$, or
		\item $T_D > k$, or
		\item $\sum\limits_{i=D}^k i \cdot t_{i} \le T_D \cdot (T_D - 1) + \sum\limits_{i=k-a}^{D-1} \min\{i, T_D\} \cdot t_{i}$.
	\end{enumerate}
\end{lemmaO}
\begin{proofO}
	Clearly, by Proposition \ref{thm:erdos-gallai}, the condition that $\sum_{i=k- a}^k i \cdot t_{i}=\sum_{j=1}^n d_j$ is even is necessary for any graphic sequence $d_1, \ldots, d_n$.
	
	We now prove that if $d_1, \ldots, d_n$ is graphic, then the third condition (hence, at least one condition) is satisfied for each $D \in [k-a,a]$.
	Note that the third condition corresponds exactly to the inequality of Erd\H{o}s and Gallai from Proposition \ref{thm:erdos-gallai} with $t=T_D$, just expressed in the terms of $t_{i}$ instead of $d_j$.
	Hence, if $d_1, d_2, \ldots, d_n$ is graphic, then the third condition is satisfied for each $D$ such that $T_D \in [n]$.
	The only case left is $T_D=0$.
	In this case, the left part of the inequality is equal to zero, so the third condition is also satisfied.
	It is left to prove the theorem statement in the other direction.
	
	By the result of Tripathi and Vijay \cite{Tripathi2003}, to check that $d_1, \ldots, d_n$ is graphic it is sufficient to check the condition of Proposition \ref{thm:erdos-gallai} for all values of $t$ such that $d_t > d_{t+1}$, and for $t=n$.
	Note that all such values of $t$ are values of $T_D$ for some choices of $D$.
	That is, if $d_t > d_{t+1}$, let $D=d_t$.
	Then $T_D=\sum_{i=D}^k t_{i}=|\{i: d_i \ge D\}|=t$.
	If we choose $D=k - a$, we achieve $T_D=n$.
	Thus, to check that $d_1, \ldots, d_n$ is graphic it is enough to check the condition of Proposition \ref{thm:erdos-gallai} for $t=T_D$, for each $D \in [k-a,k]$.
	This is exactly what the third condition checks, except for the choices of $D$ when one of the first two conditions is satisfied.
	Hence, it is now left to show that when $T_D < k-a$ or $T_D > k$, then the third condition is satisfied automatically.
	
	First, consider the case $T_D > k$.
	Note that $$\sum_{i=D}^k i \cdot t_{i} \le \sum_{i=D}^k k \cdot t_{i} = k \cdot \sum_{i=D}^k t_{i} = k \cdot T_D \le (T_D-1)\cdot T_D.$$
	Thus, the third condition is satisfied for $T_D > k$.
	
	Now consider the case $T_D < k-a$.
	Then $$\sum_{i=k-a}^{D-1} \min\{i,T_D\} \cdot t_{i}=\sum_{i=k-a}^{D-1} T_D\cdot t_{i}=T_D \cdot \sum_{i=k-a}^{D-1} t_{i}=T_D \cdot (n-T_D).$$
	As $k<n$, we have that 
	$$\sum\limits_{i=D}^k i \cdot t_{i}\le (n-1) \cdot \sum_{i=D}^k t_{i}=(n-1)\cdot T_D=T_D\cdot(T_D-1)+T_D\cdot (n-T_D),$$
	which is equivalent to the inequality of the third condition of the lemma.
	Thus, for $T_D < k-a$ the third condition is also satisfied.
	The concludes the proof of the lemma.
\end{proofO}

Lemma~\ref{lemma:check_eg} still  does not yield  directly the desired  encoding in \probILPshort.
Though $T_D$ can be expressed as a sum of $t_{i}$'s, the summand $T_D\cdot(T_D-1)$ is not allowed in a \emph{linear} equation with $T_D$ being a variable.
However, since the number of $T_D$'s is at most $|C|+1$, for each $T_D$ the algorithm can guess whether $T_D>k$, $T_D<k-|C|$ or the exact value of $T_D\in [k-|C|,k]$.
For each $T_D$ it leads to at most $|C|+3$ options, so there are at most $|C|^{\O(|C|)}$ possible options   in total.
This allows us to use 
  the values of $T_D$'s in  \probILPshort   as constants. Since 
  the variables of type $t_{i}$ are the only remaining variables, we can write the corresponding constraints as 
   linear inequalities.
\ifshort
\else
We are now ready to state the main result of this section and give it a formal proof, accumulating ideas discussed above in this section. The proof also contains the full description of the constructed linear program.
\fi
\begin{theoremO}\label{thm:vc:edgecore}
	\textsc{Edge $k$-Core} admits an \FPT~algorithm when parameterized by the vertex cover number.
	The running time of this algorithm is $2^{\O(\vc \cdot 3^{\vc})}\cdot \polyn$, where $\vc$ is the minimum size of a vertex cover of the input $n$-vertex graph.
\end{theoremO}

\begin{proofO}
	Let $(G,b,k,p)$ be an  instance of \probCore, and let  $C \subseteq V(G)$ be a minimum vertex cover in $G$.
	Denote $\vc=|C|$.
	Note that if $C$ is not given as input, it can be found in $2^{\vc} \cdot \polyn$ running time with the well-known \FPT~branching algorithm for \textsc{Vertex Cover}.
	
	Let $H \subseteq V(G)$ be a subset of the vertices of $G$ such that
	$H$ becomes a $k$-core in $G$ after adding at most $b$ edges to $G$.
	We can assume that the set of added edges $B$ is a subset of $\binom{H}{2}$, otherwise $B$ is not minimal.
	Since we are not interested in the vertices outside of $H$, let our algorithm iterate over all possible values of $H \cap C$, and delete all vertices in $C\setminus H$ from $G$ for a fixed choice of $H \cap C$.
	Clearly, there are $2^{\vc}$ choices of $H \cap C$, so this gives $2^{\vc}$ multiplier to the running time of the algorithm.
	We can now assume that we are looking for a $k$-core that contains all vertices of the vertex cover of $G$, i.e.\ $C \subseteq H$.
	Note that $C$ is a vertex cover in $G[H]$ and denote by $I=H\setminus C$ the independent set in $G[H]$.
	
	Let the algorithm also guess the edges that are added between the vertices of $C$.
	There are at most $2^{\binom{\vc}{2}}$ (at most $2^{\mathcal{O}(\vc^2)}$) possible options of adding at most $b$ of these edges.
	Our algorithm iterates over all possible options of adding edges between the vertices of $C$, adds these edges in $G$ and decreases the budget $b$ accordingly for each option.
	This adds a $2^{\mathcal{O}(\vc^2)}$ multiplier to the running time of the algorithm.
	We now assume that such guess is fixed and we can only add at most $b$ edges to $G$ such that each edge added has at least one endpoint in $I$.
	
	We are now ready to describe the ILP formulation of \textsc{Edge $k$-Core}.
	The whole description of \textsc{Edge-$k$-Core-ILP} is presented in Figure \ref{fig:ILP}.
	
	\begin{figure}
		\centering
		\begin{tabular}{r@{\hskip5pt}lr}
			\hline
			\multicolumn{3}{l}{\textsc{Edge-$k$-Core-ILP}} \\
			\hline
			\multicolumn{2}{l}{achieve} \\
			\multicolumn{3}{l}{\inlineequation[inequ:budget]{\sum\limits_{S' \subseteq C}\sum\limits_{S \subseteq S'} |S'\setminus S| \cdot y_{S,S'}+\sum\limits_{i=\max\{0,k-|C|\}}^k \sum\limits_{j=\max\{0,k-|C|-1\}}^i (i-j) \cdot u_{i,j}+z \le b}} \\[20pt]
			\multicolumn{2}{l}{subject to} & \\[10pt]
			& \inlineequation[inequ:xs]{x_S \le \left|\{v \mid N_G(v)=S,\, v \notin C\}\right|,} & $\forall$ $S \subseteq C$, \\[10pt]
			
			& \inlineequation[inequ:pvertices]{|C|+\sum\limits_{S \subseteq C} x_S \ge p,}\\[10pt]
			
			& \inlineequation[inequ:typechange]{\sum\limits_{S \subseteq S' \subseteq C} y_{S,S'} = x_S,} & $\forall$ $S \subseteq C$, \\[10pt]
			
			& \inlineequation[inequ:cdegree]{\deg_{G[C]}(v) + \sum \limits_{S' \ni v} \sum\limits_{S\subseteq S'} y_{S,S'} \ge k,} & $\forall$ $v \in C$, \\[10pt]
			
			& \inlineequation[inequ:defcount]{s_i = \sum\limits_{S' \subseteq C, k-|S'|=i} \sum \limits_{S \subseteq S'} y_{S, S'},} & $\forall$ $i \in [k-|C|, k], i\ge 0$, \\[10pt]
			
			& \inlineequation[inequ:defchange]{\sum\limits_{j=\max\{0,k-|C|-1\}}^i u_{i,j} = s_i,} & $\forall$ $i \in [k-|C|, k]$, $i \ge 0$, \\[10pt]
			
			& \inlineequation[inequ:newdef]{t_{j}=\sum\limits_{i=j}^k u_{i,j},} & $\forall$ $j \in [k-|C|-1, k], j\ge 0$, \\[10pt]
			
			& \inlineequation[inequ:bigzero]{t_{j}=0,} & $\forall$ $j \in [D_{\max}+1, k]$ \\[10pt]
			
			& \inlineequation[inequ:dmaxone]{t_{D_{\max}}>0,}\\[10pt] 
			
			& \inlineequation[inequ:dmax]{D_{\max} \le \sum\limits_{j=\max\{0,k-|C|-1\}}^k t_{j} - 1,} & \\[10pt]
			
			& \inlineequation[inequ:even]{2z = \sum\limits_{j=\max\{0,k-|C|-1\}}^{D_{\max}} j \cdot t_{j},} \\[10pt]
			
			& \inlineequation[inequ:tds]{\sum\limits_{j=D}^{D_{\max}} t_{j}=T_D,} & $\forall$ $ D \in [L, R]$, if $L\le R$,\\[10pt]
			
			& \inlineequation[inequ:right]{\sum\limits_{j=R+1}^{D_{\max}} t_{j} < k-|C|-1}, & if $D_{\max} > R$, \\[10pt]
			
			& \inlineequation[inequ:left]{\sum\limits_{j=L-1}^{D_{\max}} t_{j} > D_{\max},} & if $L > \max\{0,k-|C|-1\}$ \\[10pt]
			
			\multicolumn{3}{@{\hskip11pt}l}{\inlineequation[inequ:erdos]{\sum\limits_{j=D}^{D_{\max}} j \cdot t_{j} \le T_D (T_D - 1) + \sum\limits_{j=\max\{0,k-|C|-1\}}^{D-1} \min\{j, T_D\} \cdot t_{j},}\,\,\,\,\, $\forall$ $ D \in [L, R]$,}\\[20pt]
			
			\multicolumn{2}{l}{and}\\[5pt]
			\multicolumn{3}{l}{$x_S \ge 0,$ $y_{S,S'}\ge 0,$ $s_i \ge 0,$ $u_{i,j} \ge 0,$ $t_{j} \ge 0,$ $ z \ge 0$.}\\
			\hline
		\end{tabular}
		\caption{The ILP formulation of \probCore. $D_{\max}$, $L$, $R$ and $T_D$'s are parameters guessed by the algorithm.
			Note that equations in the formulation should be replaced with two symmetrical inequalities in order to obtain a correct linear program.}
		\label{fig:ILP}
	\end{figure}
	
	The formulation uses values $D_{\max}, L, R, T_D$, that are not variables of the linear programming, but are values that algorithm also needs to guess.
	To explain the purpose of these values and \textsc{Edge-$k$-Core-ILP} itself, we start directly with the following claim.
	
	\begin{claim}
		Let $B \subseteq \binom{V(G)}{2}\setminus E(G) \setminus \binom{C}{2}$ be a subset of at most $b$ edges, and let $G'$ denote the graph $G+B$.
		Let $H$ be the $k$-core of $G'$.
		If $C \subseteq H$ and $|H| \ge p$, then there exist integers $D_{\max}, L, R$ and $T_D$'s (for each $D \in [L,R]$), such that \textsc{Edge-$k$-Core-ILP} is feasible for this choice of $D_{\max}, L, R, T_D$.
		Moreover, all these integers are non-negative integers in $[k-|C|-1,k]$.
	\end{claim}
	\begin{claimproof}
		
		We show how to assign values to the variables of \textsc{Edge-$k$-Core-ILP} so that every inequality is satisfied.
		We also explain the purpose of each variable.
		
		For each $S \subseteq C$, assign $x_S=|\{v \in H \setminus C \mid N_G(v)=S\}|$. Thus, $x_S$ denotes the number of vertices of type $S$ in $H$. Clearly, inequalities \ref{inequ:xs} are satisfied by such values of $x_S$.
		Then  \eqref{inequ:pvertices} counts the vertices in $H$ and ensures that there are at least $p$ of them.
		This inequality is also satisfied because $|H| \ge p$.
		
		Consider now how the type of a vertex $v \in H \setminus C$ changes in $G'$.
		That is, in $N_G(v)=S$, but when edges in $B$ are added to $G$, $v$ may change its type and $N_{G'}(v)=S'$ for some $S' \supseteq S$.
		Variables $y_{S,S'}$ correspond to the amount of vertices of type $S$ in $G[H]$ that change their type to $S'$ in $G'[H]$.
		Thus, $y_{S,S'}=\{v \in H\setminus C \mid N_G(v)=S, N_{G'}(v')=S'\}$.
		Equalities \ref{inequ:typechange} ensure that these amounts agree with the number of vertices of certain type in $G[H]$, and clearly are satisfied.
		Note that this handles all edges in $B$ that have one endpoint in $I$ and the other endpoint in $C$.
		Thus, one may now evaluate the degree of each $v \in C$ by counting vertices of certain types in $G'[H]$.
		Inequalities \ref{inequ:cdegree} ensure that the degree of each $v \in C$ is at least $k$.
		Each one of these inequalities is satisfied since $H$ is a $k$-core in $G'$.
		We now assume that $B\subseteq \binom{H}{2}$, otherwise all edges with an endpoint outside of $B$ can be removed from $B$ while the $k$-core $H$ in $G'$ and $G'[H]$ remains the same.
		Hence, $G'[H]$ is exactly the graph $G[H]+B$.

		The rest of inequalities of \textsc{Edge-$k$-Core-ILP} handle edge additions inside $G[I]$.
		Divide $B$ into two parts, $B=B_C \cup B_I$, where $B_C$ is the set of all edges in $B$ having at least one endpoint in $C$, and $B_I$ is the set of edges in $B$ having both endpoints in $I$.
		After adding all edges in $B_C$ to $G[H]$ we obtain an intermediate graph $G[H]+B_C$. A vertex $v \in H \cap I$ of type $S'$ in $G[H]+B_C$ has deficiency $\defc_{G[H]+B_C}(v)=\max\{0, k-|S'|\}$.
		Then \eqref{inequ:defcount} evaluates the number of vertices with certain deficiencies in $G[H]+B_C$.
		Clearly, all possible deficiency values lie in $[k-|C|,k]$, so $s_i$ indeed denotes the number of vertices with deficiency $i$ for each non-negative $i \in [k-|C|,k]$.
		
		We now consider $B_I \subseteq \binom{I}{2}$.
		Fix an arbitrarily ordering of edges in $B_I$ and consider how edges of $B_I$ are added to $G[H]+B_C$ in this order.
		Some of these edge additions are good and some are bad.
		If there are edge additions that do not change deficiency of any vertex, we can remove such edge from $B$ and $G[H]+B$ still remains a $k$-core.
		Thus, $B_I$ can be divided into two parts $B_I=B_I^{(2)} \sqcup B_I^{(1)}$, where $B_I^{(2)}$ corresponds to the set of good edges and $B_I^{(1)}$ corresponds to the set of bad edges.
		Correspondingly, their additions change total deficiency of $G[H]+B_C$ by two and by one when some order is fixed.
		Consider the graph $G[I]+B_I^{(2)}$ formed by the good edges in $B$.
		For each $v \in I$, $\deg_{G[I]+B_I^{(2)}}(v) \le \defc_{G[H]+B_C}(v)$, as each edge in $B_{I}^{(2)}$ is good when added to $G[H]+B_C$.
		Let $t_{j}=\{v \in I \mid \deg_{G[I]+B_I^{(2)}}(v)=j\}$ be the number of vertices of degree $j$ in $G[I]+B_I^{(2)}$.
		Then $u_{i,j}$ denotes the number of vertices in $I$ with deficiency $i$ in $G[H]+B_C$ that have degree $j$ in $G[I]+B_I^{(2)}$, and $j \le i$.
		That is, $u_{i,j}=\{v \in I \mid \defc_{G[H]+B_C}(v)=i, \deg_{G[I]+B_I^{(2)}}(v)=j\}$.
		This is described by equalities \ref{inequ:defchange} and \ref{inequ:newdef}, that are satisfied by the chosen values of $t_{j}$ and $u_{i,j}$.
		We only need to show that $j$ is always an integer in $[k-|C|-1,k]$.
		If $k \le |C|$, it is true since $j \ge 0$, so we consider $k>|C|$.
		
		Note that $u_{i,j}$ actually corresponds to adding bad edges from $B_I^{(1)}$.
		If a vertex $v \in I$ has deficiency $\defc_{G[H]+B_C}(v)=i$, but is incident with only $\deg_{G[I]+B_I^{(2)}}(v)=j$ good edges, then it requires $i-j$ bad edges in $B_I^{(1)}$ incident to it to be added in $G$.
		This can be considered as follows. 
		Initially, we are given a sequence of vertex deficiencies in $I$ in $G[H]+B_C$.
		This sequence is described by the values of $s_i$: for each $i$, we have $s_i$ vertices in $I$ with deficiency $i$, so we have $s_i$ integers equal to $i$ in the sequence.
		By   \eqref{inequ:pvertices}, there are at least $p-|C|\ge k+1-|C|$ vertices in this sequence.
		If this sequence is graphic, then we do not require any bad edge addition inside $G[H]+B_C$: just pick good edges as the edges of a graph on $|I|$ vertices corresponding to this sequence.
		Thus, if the values of $s_i$ correspond to a graphic sequence, $B_I^{(1)}$ can be chosen to be empty.
		Otherwise, we need to slightly change the sequence of deficiencies using bad edge additions, and we want to minimize these bad edge additions.
		$u_{i,j}$ specifies how many integers with value $i$ in the sequence are changed to $j$.
		This corresponds to adding $i-j$ bad edges incident to each one of $u_{i,j}$ vertices with deficiency $i$.
		
		Consider the sequence of integers from $[k-|C|,k]$ that is not graphic.
		With a single operation, we are allowed to decrease some integers of this sequence by one.
		By Proposition~\ref{thm:erdos-gallai}, we should always decrease the maximum integer in the sequence by one with a single operation.
		Suppose that we repeated this operation and obtained integer $k-|C|-2$ in the sequence at some moment.
		Then at some moment all integers in the sequence were equal to $k-|C|$, and at some moment later all integers in the sequence were equal to $k-|C|-1$.
		Since at least one of $k-|C|$ and $k-|C|-1$ is even and also there are at least $k+1-|C|$ integers in the sequence, there exists a $(k-|C|)$-regular or a $(k-|C|-1)$-regular graph consisting of $n \ge k+1-|C|$ vertices.
		Thus, at least one of the sequences consisting of integers all-equal to $k-|C|$ or to $k-|C|-1$ is graphic.
		Hence, we do not need to decrease any integer below $k-|C|-1$ in order to minimize the number of operations of changing the sequence.
		
		We have shown that $j \in [k-|C|-1, k]$. What remains is  to explain the purpose of the remaining inequalities of \textsc{Edge-$k$-Core-ILP} and to show how to choose the values of $D_{\max},L,R$ and $T_D$'s for each $D \in [L,R]$ in order to satisfy them.
		The actual purpose of inequalities \eqref{inequ:bigzero}--\eqref{inequ:erdos} is to check that $t_{j}$ corresponds to a graphic sequence according to Lemma \ref{lemma:check_eg}.
		The value of $a$ in Lemma \ref{lemma:check_eg} is chosen as $a=\min\{k,|C|+1\}$.
		Pick $D_{\max}=\max\{j: t_{j}>0\}$.
		Then \eqref{inequ:bigzero} and \eqref{inequ:dmaxone} ensure  that $D_{\max}$ is chosen correctly. 
		Inequality \eqref{inequ:dmax} guarantees that there are at most $D_{\max}+1$ integers in the sequence, since otherwise it is not graphic.
		Equality \eqref{inequ:even} checks that the sum of integers in the sequence is even.
		Note that $z=|B_I^{(2)}|$ equals the number of good edges in $B$.
		
		Then, $L$ and $R$ are the bounds of $D$ for which the third condition of Lemma \ref{lemma:check_eg} should be verified.
		Sequence $T_{\max\{0,k-|C|-1\}}, \ldots, T_{k-1}, T_{k}$ is non-increasing, so $L$ is the smallest integer for which $T_L \le k$.
		Analogously, $R$ is the largest integer for which $T_R \le k-a$.
		If $i < L$ or $i > R$, then at least one of the first two conditions of Lemma \ref{lemma:check_eg} is satisfied, and we may not check the Erd\H{o}s-Gallai inequality for this $T_i$.
		Note that it may be the case that $L>R$, then we do not  have to check the third condition of Lemma \ref{lemma:check_eg} at all.
		Otherwise, for each $D \in [L,R]$, $T_D$ is a non-negative integer in $[k-|C|-1,k]$.
		Then \eqref{inequ:tds}--\eqref{inequ:left} ensure that $L,R$ and $T_D$'s are chosen correspondingly to the values of $t_{j}$ and are satisfied for our choices of these parameters.
		Finally,  \eqref{inequ:erdos} is responsible  that the third condition of Lemma \ref{lemma:check_eg} holds for each $D \in [L,R]$.
		Thus, if $t_{j}$ satisfies \eqref{inequ:bigzero}--\eqref{inequ:erdos} for the correct choice of $D_{\max},L,R$ and $T_D$'s, then by Lemma~\ref{lemma:check_eg}, $t_{j}$ corresponds to a graphic sequence.
		Note that for each choice of $t_{j}$ there is exactly one feasible choice of the parameters $T_D, D_{\max}, L$, and $R$.
		
		It is left to explain the role of  \eqref{inequ:budget} of \textsc{Edge-$k$-Core-ILP}.
		We show that the left part of this inequality equals $|B|$.
		Note that $|B_C|=\sum\limits_{S' \subseteq C}\sum\limits_{S \subseteq S'} |S'\setminus S| \cdot y_{S,S'}$, as edges in $B_C$ can be counted using the number of type changes from $S$ to $S'$, and a type change from $S$ to $S'$ is done with $|S' \setminus S|$ edge additions.
		The correspondence between bad edges in $B_I$ and the values $u_{i,j}$ is described above, so the second summand in inequality \ref{inequ:budget} corresponds to $|B_I^{(1)}|$.
		Finally, the summand $z$ equals $|E(G_I)|=|B_I^{(2)}|$.
		Thus, the left part of the inequality equals $|B_C|+|B_I^{(1)}|+|B_I^{(2)}|=|B|$, and $|B| \le b$.
		The proof of the claim is completed.
	\end{claimproof}
	
	We have shown that if $G$ can be completed to a graph with a $k$-core of size at least $p$, then \textsc{Edge-$k$-Core-ILP} is feasible for the correct choice of the parameters.
	Note that each of $D_{\max}, L, R$ and $T_D$'s are integers in $[k-|C|-1,k]$, so there are $|C|+2$ options for each of them.
	Since there are at most $R-L+1 \le |C|+2$ choices of index $D$ for $T_D$, there are at most $|C|+5$ integers to choose from $[k-|C|-1,k]$.
	Clearly, this leads to a total of $\O(|C|^{|C|})$ possible options for the parameters.
	The algorithm iterates over all possible choices of these values, constructs an ILP formulation for each fixed choice and then employs the algorithm of Proposition \ref{thm:ilpalgo} to check if \textsc{Edge-$k$-Core-ILP} is feasible.
	There are $\O(3^{|C|})$ variables and inequalities in the formulation (with $y_{S,S'}$ being the majority of the variables), so the feasibility of the formulation is checked in $\O(3^{|C|})^{\O(3^{|C|})}\cdot n^{\O(1)}=3^{\O(|C| \cdot 3^{|C|})}\cdot n^{\O(1)}$.
	The outer guesses of $H \cap C$, $B \cap \binom{C}{2}$ and the parameters of ILP gives $|C|^{|C|} \cdot n^{\O(1)}$ multiplier to the running time of the algorithm.
	This multiplier is dominated by the double exponent, so the total running time is still $3^{\O(|C| \cdot 3^{|C|})}\cdot n^{\O(1)}$.
	It is only left to say that if the algorithm finds \textsc{Edge-$k$-Core-ILP} feasible for some choice of parameters, then the initial instance is a yes-instance of \probCore.
	
	\begin{claim}
		If \textsc{Edge-$k$-Core-ILP} is feasible for some choice of $D_{\max}, L, R, T_D$, then $(G,b,k,p)$ is a yes-instance of \probCore.		
	\end{claim}
	\begin{claimproof}
		
		One needs to follow the proof of the previous claim.
		It can be shown that from a feasible solution to \textsc{Edge-$k$-Core-ILP} can be constructed a set of vertices $H \subseteq V(G)$ (following the values of $x_S$) and sets of edges $B_C$ (following the values of $y_{S,S'}$), $B_I^{(1)}$ (following the values of $u_{i,j}$) and $B_I^{(2)}$ (following that the values of $t_{j}$ corresponds to a graphic sequence).
		Then $G[H]$ can be completed to a graph of minimum degree $k$ by adding edges from $B=B_C \cup B_I^{(1)} \cup B_I^{(2)}$ to $G$.
		Inequalities ensure that $|H|\ge p$ and $|B| \le b$.
		Note that the constructed $H$ is not necessarily the $k$-core of $G'$, but it is necessarily a subset of its vertices.
	\end{claimproof}
	
	The proof is complete.
\end{proofO}


\include*{vertex-cover-lb}

	\section{Treewidth}\label{sec:treewidth}

In this section, we give an \FPT-algorithm for \textsc{Edge $k$-Core} parameterized by $\tw + k$. This improves upon the following result of Chitnis and Talmon, and we also use their algorithm as a subroutine.
\begin{proposition}[\cite{Chitnis2018}]
    \textsc{Edge $k$-Core} can be solved in time $(k + \tw)^{\O(\tw + b)} \cdot \polyn$.
    \label{prop:oldfpt}
\end{proposition}

We start with the central combinatorial result of this section which allows the algorithmic improvement.
Namely, we show that whenever the total deficiency of a graph $G$ exceeds a polynomial in $k$, $G$ can be completed to a graph of minimum degree $k$ using the minimum possible number of edges.
Also, the required edge additions can be identified in polynomial time.

We believe that this result is interesting on its own, since it considerably simplifies the problem whenever the budget is sufficiently high compared to $k$. If we are trying to identify the best vertex set $H$ which induces a $k$-core, we have to only care about the total deficiency of $G[H]$, and not of any particular structure on it.

\begin{theorem}\label{lemma:largeopt}
	For any integer $k\ge 2$, any graph $G$ with $\defc(G)\ge 3k^3$ can be completed to a graph of minimum degree $k$ using $\lceil \frac{1}{2} \defc(G) \rceil$ edge additions with a polynomial-time algorithm.
\end{theorem}
\begin{proof}
	It is enough to prove that we can satisfy all deficiencies by adding only good edges, except if $\defc(G)$ is odd, exactly one edge addition is bad.
	
	We constructively obtain a graph $G'$ of form $G+B$, initially $B = \emptyset$. The construction is a polynomial time algorithm.
	
	First, we exhaustively apply the following rule, which always does one good edge addition. If there are two distinct vertices $u, v \in V(G')$ such that $\defc_{G'}(u) > 0$, $\defc_{G'}(v) > 0$, and $uv \notin E(G')$, then add the edge $uv$ to $B$.
	Assume that the rule is no longer applicable.
	Let us denote $C = \{v \in V(G) | \defc_{G'}(v) > 0\}$, by the conditions of the rule, $C$ induces a clique in $G'$. Then, $|C| \le k$, since otherwise vertices in $C$ could not have positive deficiency.
	
	Now we exhaustively apply the new rule. Fix two vertices $u, v \in C$, such that either $u$ and $v$ are distinct, or $u = v$ and $\defc_{G'}(u) \ge 2$. Then find two distinct vertices $u', v' \in V(G') \setminus C$ such that
	$u'v' \in B$ and $uu', vv' \notin E(G')$. Since $u'v'$ is in $B$, $u'$ and $v'$ have degree exactly $k$, as previously we have only added good edges and $u', v' \notin C$.
	Delete $u'v'$ from $B$, now $u'$ and $v'$ have positive deficiencies. Add edges $uu'$ and $vv'$ to $B$, by the choice of $u'$ and $v'$ these edges are not in $E(G')$, and also both these additions are good.
	
	
	We claim that when the new rule is no longer applicable, the size of $C$ is at most one, and $\defc(G')$ is also at most one. Suppose it is not true, in this case there is always a proper choice of $u, v \in C$. Then there are no $u', v' \in V(G) \setminus C$ satisfying the conditions above. Then each edge $u'v' \in B$ is of one of the following kinds:
	\begin{enumerate}
		\item $u', v' \in C$, since $|C| \le k$, there are at most $\binom{k}{2}$ such edges;
		\item one of $u'$, $v'$ is in $C$ and the other is not in $C$, there are at most $k(k - 1)$ edges of this kind, since $|C| \le k$ and degrees in $C$ are less than $k$;
		\item $u', v' \notin C$, and either $uu' \in E(G')$ or $vv' \in E(G')$; there are at most $k(k - 1)$ vertices adjacent to $C$, and each of them has at most $k$ incident edges from $B$, so there are at most $k^2(k - 1)$ such edges.
	\end{enumerate}
	Then the size of $B$ is at most $\binom{k}{2} + k(k - 1) + k^2(k-1) < 2k^3$. However, $\defc(G) = 2|B| + \defc(G')$, and $\defc(G') \le |C| \cdot k  \le k^2$. So $\defc(G) < 3k^3$ contradicting the statement.
	
	Therefore, by the constructed sequence of good additions we reached the situation when $|C|$ and $\defc(G')$ are both at most one. If $C$ is empty, we are done. If $C$ consists of one vertex $u$, then its deficiency is one. Since $\defc(G) = 2|B| + \defc(G')$, $\defc(G)$ is odd, and we have one more edge addition. Then we add to $B$ an edge from $u$ to any other vertex $v$ such that $uv \notin E(G')$; this is always possible since $\deg_{G'}(u) < k$, and $V(G) > k$ because $\defc(G) \ge 3k^3$.
\end{proof}

The intuition to our \FPT~algorithm is as follows. When we can obtain a sufficiently large $k$-core by adding a number of edges $b<3k^3$, the algorithm from Proposition~\ref{prop:oldfpt} suffices. Otherwise $b\ge 3k^3$ and by Theorem~\ref{lemma:largeopt} we can focus on finding a vertex subset in $G$ of size at least $p$ minimizing the total deficiency of the induced subgraph.
We show how to do that with a dynamic programming over a tree decomposition.
\ifshort
\else
First we introduce some preliminaries about treewidth and tree decompositions from \cite{cygan2015parameterized}.
A {\em tree decomposition} of
a graph $G$ is a pair $\mathcal{T}=(T,\{X_t\}_{t\in V(T)})$, where $T$ is a tree whose every node $t$ is assigned a vertex subset $X_t\subseteq V(G)$, called a bag,
such that the following three conditions hold:
\begin{description}
\item[(T1)] $\bigcup_{t\in V(T)} X_t =V(G)$.
\item[(T2)] For every $uv\in E(G)$, there exists a node $t$ of $T$ such that bag $X_t$ contains both $u$ and $v$.
\item[(T3)] For every  $u\in V(G)$, the set $T_u = \{t\in V(T) : u\in X_t\}$, i.e., the set of nodes whose corresponding  bags contain $u$, induces
a connected subtree of $T$.
\end{description}
The {\em width} of tree decomposition $\mathcal{T}=(T,\{X_t\}_{t\in V(T)})$  equals $\max_{t\in V(T)} |X_t| - 1$, that is, the maximum size of its bag minus $1$. The {{\em
treewidth}} of a graph $G$, denoted by $\tw(G)$, is the minimum possible width of a tree decomposition of $G$.

For the simplicity we will only consider so-called {\em nice tree decompositions}.
A tree decomposition $(T,\{X_t\}_{t\in V(T)})$ rooted at $r \in T$ is {\em{nice}} if the following conditions are satisfied:
\begin{itemize}
\item $X_r=\emptyset$ and $X_\ell=\emptyset$ for every leaf $\ell$ of $T$. In other words, all the leaves as well as the root contain empty bags. 
\item Every non-leaf node of $T$ is of one of the following three types:
\begin{itemize}
  \item{\bf{Introduce node}}: a node $t$ with exactly one child $t'$
    such that $X_t=X_{t'}\cup\{v\}$ for some vertex $v\notin X_{t'}$; we say
    that $v$ is {\emph{introduced}} at $t$.
  \item{\bf{Forget node}}: a node $t$ with exactly one child $t'$ such
    that $X_t=X_{t'}\setminus \{w\}$ for some vertex $w\in X_{t'}$; we say
    that $w$ is {\emph{forgotten}} at $t$.
  \item{\bf{Join node}}: a node $t$ with two children $t_1,t_2$ such
    that $X_t=X_{t_1}=X_{t_2}$.
\end{itemize}
\end{itemize}
As shown in \cite{cygan2015parameterized}, any tree decomposition could be turned into a nice one in polynomial time. For a node $t$ in a nice tree decomposition $\mathcal{T}=(T,\{X_t\}_{t\in V(T)})$ denote by $V_t$ the union of all the bags present in the subtree of $T$ rooted at $t$, including $X_t$.
\begin{proposition}[\cite{cygan2015parameterized}]
    Given a graph $G$ of treewidth $\tw$, a nice tree decomposition of an $n$-vertex graph $G$ of width $4\tw + 4$ can be constructed in time $8^\tw \cdot \polyn$.
    \label{prop:computing_td}
\end{proposition}

Now we are ready to show our algorithm.
\fi

\newcommand{\hS}{\widehat{S}}
\begin{lemmaO}
    Given an $n$-vertex graph $G$ of treewidth $\tw$ and integers $k$, $p$, the value
    \[\min \{\defc(G[\hS]) : \hS \subseteq V(G), |\hS| \ge p\}\]
    can be computed in time $k^{\O(\tw)} \cdot \polyn$.
    \label{lemma:deficiency}
\end{lemmaO}
\ifshort
\else
\begin{proof}
    Consider a nice tree decomposition $\mathcal{T}=(T,\{X_t\}_{t\in V(T)})$ of $G$ given by algorithm in Proposition~\ref{prop:computing_td}, width of $\mathcal{T}$ is $\O(\tw)$. For every node $t$, every $S \subset X_t$, every function $f : S \to \{0, \cdots, k\}$, and every number $s \in \{0, \cdots, |V_t|\}$ define the following value:
    \[c[t, S, f, s] = \min \{\defc(G[\hS]) : \hS \in \mathcal{C}[t, S, f, s]\},\]
    where
    \[\mathcal{C}[t, S, f, s] = \{\hS \subset V_t : \hS \cap X_t = S, |\hS| = s, \text{ and for every } v \in S, \defc_{G[\hS]}(v) = f(v)\}.\]
    Next, we show how the values of $c[\cdot, \cdot, \cdot, \cdot]$ are computed for all kinds of nice tree decomposition nodes.

    \textbf{Leaf node.} If $t$ is a leaf node, there is only one value $c[t, \emptyset, f_\emptyset, 0] = 0$, where $f_\emptyset$ denotes the empty function to $\{0, \cdots, k\}$.

    \textbf{Introduce node.} Let $t$ be an introduce node, $t'$ be its only child and $v$ be the vertex such that $v \notin X_{t'}$, $X_t = X_{t'} \cup \{v\}$.
    We claim that for any $S$, $f$, $s$ from the definition of $c[t, \cdot, \cdot, \cdot]$
    \begin{equation}
    c[t, S, f, s] = \begin{cases}
            c(t', S, f, s) & \text{ if } v \notin S;\\
            c(t', S \setminus \{v\}, f|_{S \setminus \{v\}}, s - 1) + f(v) & \text{ if } v \in S, f(v) = \defc_{G[S]}(v), s > 0;\\
            \infty & \text{ otherwise }.
\end{cases}
        \label{eqn:dp_introduce}
    \end{equation}
    If $v \notin S$, \eqref{eqn:dp_introduce} clearly holds since the families of the sets $\hS$ are the same in both parts of the equation. If $v \in S$, then for $\hS \in \mathcal{C}[t, S, f, s]$ to exist $s$ must be greater than zero since $v \in \hS$. Also, $f(v)$ must be equal to $\defc_{G[S]}(v)$ since
    there are no edges from $v$ to $V_t \setminus X_t$ by the definition of a nice tree decomposition, and thus $\defc_{G[S]}(v) = \defc_{G[\hS]}(v)$. This shows that the third case of \eqref{eqn:dp_introduce} holds, so only the second case remains.

    When the conditions of the second case hold, we show that for each $\hS$ in $\mathcal{C}[t, S, f, s]$ there is a corresponding $\hS'$ in $\mathcal{C}(t', S \setminus \{v\}, f|_{S \setminus \{v\}}, s - 1)$, this is a one-to-one correspondence, and $\defc(G[\hS]) = \defc(G[\hS']) + \defc_{G[S]}(v)$. Namely, $\hS' = \hS \setminus \{v\}$, and this concludes the proof of \eqref{eqn:dp_introduce}.

    \textbf{Forget node.} Let $t$ be a forget node, $t'$ be its only child and $v$ be the vertex such that $v \notin X_{t}$, $X_{t'} = X_t \cup \{v\}$.
    We claim that for any $S$, $f$, $s$ from the definition of $c[t, \cdot, \cdot, \cdot]$
    \begin{equation}
        c[t, S, f, s] = \min \{c(t', S, f, s)\} \cup \{c(t', S \cup \{v\}, f', s) : f'|_S = f\}.
        \label{eqn:dp_forget}
    \end{equation}
    First, consider any $\hS \in \mathcal{C}[t, S, f, s]$. If $v \notin \hS$, then $\hS$ is also in $\mathcal{C}(t', S, f, s)$. If $v \in \hS$, then $\hS$ is in $\mathcal{C}[t', S \cup \{v\}, f', s + 1]$, where $f'|_S = f$ and $f(v) = \defc_{G[\hS]}(v)$.

    In the other direction, if $\hS \in \mathcal{C}[t', S, f, s]$ or $\hS \in \mathcal{C}[t', S \cup \{v\}, f', s]$ where $f'|_S = f$, then $\hS \in \mathcal{C}[t, S, f, s]$.

    \textbf{Join node.} Let $t$ be a join node, $t_1$ and $t_2$ be its children, such that $X_t = X_{t_1} = X_{t_2}$.
    We claim that for any $S$, $f$, $s$ from the definition of $c[t, \cdot, \cdot, \cdot]$
    \begin{multline}
        c[t, S, f, s] = \min_{s_1 + s_2 = s + |S|, f_1, f_2} \left(c(t_1, S, f_1, s_1) +\vphantom{\sum_{v \in S}}\right.\\
        \left.c(t_2, S, f_2, s_2) + \sum_{v \in S} (f(v) - f_1(v) - f_2(v))\right),
        \label{eqn:dp_join}
    \end{multline}
    where $f_1$ and $f_2$ are such that for every $v \in S$, $f(v) = \max(0, f_1(v) + f_2(v) - k - \deg_{G[S]}(v))$. To show the correctness of \eqref{eqn:dp_join}, first we show that the right side of \eqref{eqn:dp_join} is at most the left side. Consider an $\hS \in \mathcal{C}[t, S, f, s]$ on which the minimum is attained, i.e. $c[t, S, f, s] = \defc(G[\hS])$. Denote $\hS_1 = \hS \cap V_{t_1}$, $\hS_2 = \hS \cap V_{t_2}$, $s_1 = |\hS_1|$, $s_2 = |\hS_2|$. Since $\hS_1 \cap \hS_2 = \hS \cap V_{t_1} \cap V_{t_2} = \hS \cap X_t = S$, $s_1 + s_2 = s + |S|$ holds.
    Define also $f_1$ and $f_2$ as $f_1(v) = \defc_{G[\hS_1]}(v)$, $f_2(v) = \defc_{G[\hS_2]}(v)$. By definition, $\hS_1 \in \mathcal{C}[t_1, S, f_1, s_1]$ and $\hS_2 \in \mathcal{C}[t_2, S, f_2, s_2]$. Since for every $v \in S$
    \begin{multline}
    \defc_{G[\hS]}(v) = \max(0, k - \deg_{G[\hS]}(v)) =\\
    \max(0, k - \deg_{G[\hS_1]}(v) - \deg_{G[\hS_2]}(v) + \deg_{G[S]}(v)) =\\
    \max(0, \defc_{G[\hS_1]}(v) + \defc_{G[\hS_2]}(v) - \deg_{G[S]}(v)) =\\
    \max(0, f_1(v) + f_2(v) - \deg_{G[S]}(v)),
    \label{eqn:dp_join_f}
    \end{multline}
    and $f(v) = \defc_{G[\hS]}(v)$, the condition on $f$, $f_1$ and $f_2$ holds. Thus, the right side of \eqref{eqn:dp_join} is at most
    \begin{multline}
        \defc(G[\hS_1]) + \defc(G[\hS_2]) + \sum_{v \in S} (f(v) - f_1(v) - f_2(v)) =\\
    \defc(G[\hS_1]) - \sum_{v \in S} \defc_{G[\hS_1]}(v) + \defc(G[\hS_2]) - \sum_{v \in S} \defc_{G[\hS_2]}(v) + \sum_{v \in S} \defc_{G[\hS]}(v) =\\
        \sum_{v \in \hS_1 \setminus S} \defc_{G[\hS_1]} (v) + \sum_{v \in \hS_2 \setminus S} \defc_{G[\hS_2]} (v) + \sum_{v \in S} \defc_{G[\hS]}(v) = \\
        \sum_{v \in \hS} \defc_{G[\hS]} (v) = \defc(G[\hS]),
        \label{eqn:dp_join_value}
    \end{multline}
    since $(\hS_1 \setminus S) \uplus (\hS_2 \setminus S) \uplus S = \hS$, and for every $v \in \hS_1 \setminus S$, $\defc_{G[\hS_1]}(v) = \defc_{G[\hS]}(v)$ as $v$ is adjacent only to vertices in $\hS_1$, the analogous holds for every $v \in \hS_2 \setminus S$. Since $\defc(G[\hS]) = c[t, S, f, s]$, the first part of the correctness proof is done.

    Second, we show that the left side of \eqref{eqn:dp_join} is at most the right side of \eqref{eqn:dp_join}. Consider $s_1$, $s_2$, $f_1$, $f_2$ on which the minimum of the left side of \eqref{eqn:dp_join} is attained, and consider also $\hS_1 \in \mathcal{C}[t_1, S, f_1, s_1]$ and $\hS_2 \in \mathcal{C}[t_2, S, f_2, t_2]$ such that $\defc(G[\hS_1]) = c[t_1, S, f_1, s_1]$ and $\defc(G[\hS_2]) = c[t_2, S, f_2, s_2]$. Let $\hS = \hS_1 \cup \hS_2$, we claim that $\hS \in \mathcal{C}[t, S, f, s]$.
    Since $X_t = X_{t_1} = X_{t_2}$ and $\hS_1 \cap X_{t_1} = \hS_2 \cap X_{t_2} = S$, $\hS \cap X_t$ is also $S$. Since $\hS_1 \cap \hS_2 = S$, $|\hS| = |\hS_1| + |\hS_2| - |S| = s_1 + s_2 - |S| = s$. The equation \eqref{eqn:dp_join_f} holds with the new choice of $\hS$, $\hS_1$, and $\hS_2$; thus, $\defc_{G[\hS]}(v) = f(v)$ for every $v \in S$, and $\hS \in \mathcal{C}[t, S, f, s]$. Finally, \eqref{eqn:dp_join_value} also holds, and $\defc(G[\hS])$ is equal to the value of the right side of \eqref{eqn:dp_join}. Thus the correctness of \eqref{eqn:dp_join} is fully proven.

    We compute all values $c[\cdot, \cdot, \cdot, \cdot]$ by going through nodes of $T$ in the tree order, starting from the leaves, and applying one of the above formulas, depending on the type of node $t$, for all possible values of $S$, $f$, $s$ at this node. Finally, at the root node $r$ for every number $s \in \{0, \cdots, n\}$ we have the value $c[r, \emptyset, f_\emptyset, s]$, which is equal to $\min_{\hS \subset V(G), |\hS| = s} \defc(G[\hS])$. Finally, we return the value $\min_{p \le s \le n} c[r, \emptyset, f_\emptyset, s]$.

    For each of the $\O(k n)$ nodes of $t$, there are $2^{|X_t|}$ variants of $S$, at most $(k + 1)^{|X_t|}$ variants of $f$, and $n + 1$ variants of $s$ in the definition of $c[t, S, f, s]$. So the total number of values computed is $k^{\O(\tw)} \cdot \polyn$, and the total running time is also $k^{\O(\tw)} \cdot \polyn$.
\end{proof}
\fi

\ifshort
\else
With all pieces together, we are ready to prove the main algorithmic result of this section.
\fi

\begin{theoremO}
	\textsc{Edge $k$-Core} admits an \FPT~algorithm when parameterized by the combined parameter $\tw+k$.
\end{theoremO}
\begin{proofO}
    Consider the instance $(G, b, k, p)$ of \textsc{Edge $k$-Core}. Assume $p > k$ since the size of a $k$-core is always at least $k + 1$.

    If $b \le 3k^3$, we run the algorithm of Chitnis and Talmon given by Proposition~\ref{prop:oldfpt} and return its result. Otherwise, we run the algorithm from Lemma~\ref{lemma:deficiency} on the input $(G, k, p)$, and obtain the value $d$ such that there is a $\hS^* \subset V(G)$ of size at least $p$ with $\defc(G[\hS^*]) = d$, and $\defc(G[\hS]) \ge d$ for every $\hS \subseteq V(G)$ of size at least $p$.

    Now, if $d < 3k^3$, we report a yes-instance, since $b \ge 3k^3$, and $G[\hS^*]$ can be trivially turned into a $k$-core. Namely, for each $v \in \hS^*$, while $\deg_{G[\hS^*]} < k$ add an edge connecting $v$ and some other vertex in $\hS^*$ arbitrarily. This is always possible since $|\hS^*| > k$, and the number of edges added is at most $d$.

    If $d \ge 3k^3$, we compare $b$ to $\lceil \frac{1}{2} d\rceil$. If $b \ge \lceil \frac{1}{2} d \rceil$, we report a yes-instance, since $G[\hS^*]$ can be turned into a $k$-core using $\lceil \frac{1}{2} \defc (G[\hS^*]) \rceil$ edge additions by Lemma~\ref{lemma:largeopt}. If $b < \lceil \frac{1}{2} d \rceil$, we report a no-instance, since for every $\hS \subset V(G)$ completing $G[\hS]$ to a $k$-core requires at least $\lceil \frac{1}{2} \defc(G[\hS]) \rceil$ edge additions, and $d \le \defc(G[\hS])$ for every $\hS$ of size at least $p$.

    It remains to bound the running time. In the case $b \le 3k^3$, the running time is $(k + \tw)^{\O(k^3 + \tw)} \cdot \polyn$ by Proposition~\ref{prop:oldfpt}. In the case $b \ge 3k^3$ the algorithm from Lemma~\ref{lemma:deficiency} dominates the running time, which is $k^{\O(\tw)} \cdot \polyn$. Thus, the whole algorithm is FPT parameterized by $\tw + k$.

\end{proofO}

	\bibliographystyle{plainurl}
	\bibliography{ref}
\end{document}